\documentclass[twocolumn]{aastex63}
\usepackage{natbib}
\newcommand{\squiggle}{SQuIGG$\vec{L}$E \,}
\newcommand{\squigglecomma}{SQuIGG$\vec{L}$E}

\newcommand{\hdelta}{$H_{\delta,A}$ \,}
\newcommand{\hdeltacomma}{$H_{\delta,A}$}

\newcommand{\fonegyr}{$f_{\mathrm{1 Gyr}}$\,}
\newcommand{\fonegyrcomma}{$f_{\mathrm{1 Gyr}}$}

\newcommand{\logM}{log($M_\star/M_\odot$) \,}

\newcommand{\deltasfr}{$\Delta\mathrm{SFR}$\,}
\newcommand{\deltasfrcomma}{$\Delta\mathrm{SFR}$}

\newcommand{\nhdeltafour}{$1727$ \,}
\newcommand{\nhdeltafive}{$1035$ \,}
\newcommand{\nsquiggle}{$324$ \,}

\newcommand{\nlrgtot}{$17217$ \,}

\usepackage{amsmath} 
\newcommand{\angstrom}{\text{\normalfont\AA }{ }}
\newcommand{\angstromcomma}{\text{\normalfont\AA }{}}

\usepackage{totcount}
\newtotcounter{citnum} 
\def\oldbibitem{} \let\oldbibitem=\bibitem
\def\bibitem{\stepcounter{citnum}\oldbibitem}

\usepackage{rotating}

\usepackage{float}
\usepackage{textcomp}
\usepackage{makecell}

\date{\today}

\shorttitle{Recently Quenched Galaxies in DESI}

\begin{document}

\title{DESI Survey Validation Spectra Reveal an Increasing Fraction of Recently Quenched Galaxies at $z\sim1$}

\author[0000-0003-4075-7393]{David J. Setton}
\affiliation{Department of Physics and Astronomy and PITT PACC, University of Pittsburgh, Pittsburgh, PA 15260, USA}

\author[0000-0002-5665-7912]{Biprateep Dey}
\affiliation{Department of Physics and Astronomy and PITT PACC, University of Pittsburgh, Pittsburgh, PA 15260, USA}

\author[0000-0002-3475-7648]{Gourav Khullar}
\affiliation{Department of Physics and Astronomy and PITT PACC, University of Pittsburgh, Pittsburgh, PA 15260, USA}

\author[0000-0001-5063-8254]{Rachel Bezanson}
\affiliation{Department of Physics and Astronomy and PITT PACC, University of Pittsburgh, Pittsburgh, PA 15260, USA}

\author[0000-0001-8684-2222]{Jeffrey A. Newman}
\affiliation{Department of Physics and Astronomy and PITT PACC, University of Pittsburgh, Pittsburgh, PA 15260, USA}

\author[0000-0003-0822-452X]{Jessica N. Aguilar}
\affiliation{Lawrence Berkeley National Laboratory, 1 Cyclotron Road, Berkeley, CA 94720, USA}

\author[0000-0001-6098-7247]{Steven Ahlen}
\affiliation{Boston University, 590 Commonwealth Avenue, Boston, MA 02215, USA}

\author[0000-0001-8085-5890]{Brett H. Andrews}
\affiliation{Department of Physics and Astronomy and PITT PACC, University of Pittsburgh, Pittsburgh, PA 15260, USA}

\author{David Brooks}
\affiliation{Department of Physics \& Astronomy, University College London, Gower Street, London, WC1E 6BT, UK}

\author{Axel de la Macorra}
\affiliation{Instituto de F\'{\i}sica, Universidad Nacional Aut\'{o}noma de M\'{e}xico,  Cd. de M\'{e}xico  C.P. 04510,  M\'{e}xico}

\author[0000-0002-4928-4003]{Arjun~Dey}
\affiliation{NSF's NOIRLab, 950 N. Cherry Avenue, Tucson, AZ 85719, USA}

\author[0000-0002-8281-8388]{Sarah Eftekharzadeh}
\affiliation{SOFIA Science Center, NASA Ames Research Center, Moffett Field, CA 94035, USA}

\author[0000-0002-3033-7312]{Andreu Font-Ribera}
\affiliation{Institut de F\'{i}sica d’Altes Energies (IFAE), The Barcelona Institute of Science and Technology, Campus UAB, 08193 Bellaterra Barcelona, Spain}

\author[0000-0003-3142-233X]{Satya Gontcho A Gontcho}
\affiliation{Lawrence Berkeley National Laboratory, 1 Cyclotron Road, Berkeley, CA 94720, USA}

\author[0000-0001-6356-7424]{Anthony~Kremin}
\affiliation{Lawrence Berkeley National Laboratory, 1 Cyclotron Road, Berkeley, CA 94720, USA}

\author[0000-0002-0000-2394]{Stephanie Juneau}
\affiliation{NSF's NOIRLab, 950 N. Cherry Avenue, Tucson, AZ 85719, USA}

\author[0000-0003-1838-8528]{Martin Landriau}
\affiliation{Lawrence Berkeley National Laboratory, 1 Cyclotron Road, Berkeley, CA 94720, USA}

\author[0000-0002-1125-7384]{Aaron Meisner}
\affiliation{NSF's NOIRLab, 950 N. Cherry Ave., Tucson, AZ 85719, USA}

\author{Ramon Miquel}
\affiliation{Instituci\'{o} Catalana de Recerca i Estudis Avan\c{c}ats, Passeig de Llu\'{\i}s Companys, 23, 08010 Barcelona, Spain}
\affiliation{Institut de F\'{i}sica d’Altes Energies (IFAE), The Barcelona Institute of Science and Technology, Campus UAB, 08193 Bellaterra Barcelona, Spain}

\author[0000-0002-2733-4559]{John Moustakas}
\affiliation{Department of Physics and Astronomy, Siena College, 515 Loudon Road, Loudonville, NY 12211}

\author[0000-0001-9820-9619]{Alan Pearl}
\affiliation{Department of Physics and Astronomy and PITT PACC, University of Pittsburgh, Pittsburgh, PA 15260, USA}

\author[0000-0001-7145-8674]{Francisco Prada}
\affiliation{Instituto de Astrof\'{i}sica de Andaluc\'{i}a (CSIC), Glorieta de la Astronom\'{i}a, s/n, E-18008 Granada, Spain}

\author[0000-0003-1704-0781]{Gregory Tarlé}
\affiliation{University of Michigan, Ann Arbor, MI 48109}

\author[0000-0002-2949-2155]{Małgorzata Siudek}
\affiliation{Institute of Space Sciences (ICE, CSIC), Campus UAB, Carrer de Magrans, 08193 Barcelona, Spain}
\affiliation{Institut de Física d’Altes Energies (IFAE), The Barcelona Institute of Science and Technology, 08193 Bellaterra (Barcelona), Spain}

\author{Benjamin Alan Weaver}
\affiliation{NSF's NOIRLab, 950 N. Cherry Ave., Tucson, AZ 85719, USA}

\author[0000-0002-4135-0977]{Zhimin Zhou}
\affiliation{National Astronomical Observatories, Chinese Academy of Sciences, A20 Datun Rd., Chaoyang District, Beijing, 100012, P.R. China}

\author[0000-0002-6684-3997]{Hu Zou}
\affiliation{National Astronomical Observatories, Chinese Academy of Sciences, A20 Datun Rd., Chaoyang District, Beijing, 100012, P.R. China}

\correspondingauthor{David J. Setton}
\email{davidsetton@pitt.edu}

\submitjournal{ApJ Letters}

\begin{abstract}

We utilize $\sim17000$ bright Luminous Red Galaxies (LRGs) from the novel Dark Energy Spectroscopic Instrument Survey Validation spectroscopic sample, leveraging its deep ($\sim2.5$ hour/galaxy exposure time) spectra to characterize the contribution of recently quenched galaxies to the massive galaxy population at $0.4<z<1.3$. We use \texttt{Prospector} to infer non-parametric star formation histories and identify a significant population of recently quenched galaxies that have joined the quiescent population within the past $\sim1$ Gyr. The highest redshift subset (277 at $z>1$) of our sample of recently quenched galaxies represents the largest spectroscopic sample of post-starburst galaxies at that epoch. At $0.4<z<0.8$, we measure the number density of quiescent LRGs, finding that recently quenched galaxies constitute a growing fraction of the massive galaxy population with increasing lookback time. Finally, we quantify the importance of this population amongst massive (\logM$>11.2$) LRGs by measuring the fraction of stellar mass each galaxy formed in the Gyr before observation, \fonegyrcomma. Although galaxies with \fonegyr$>0.1$ are rare at $z\sim0.4$ ($\lesssim 0.5\%$ of the population), by $z\sim0.8$ they constitute $\sim3\%$ of massive galaxies. Relaxing this threshold, we find that galaxies with $f_\mathrm{1 Gyr}>5\%$ constitute $\sim10\%$ of the massive galaxy population at $z\sim0.8$. We also identify a small but significant sample of galaxies at $z=1.1-1.3$ that formed with \fonegyr$>50\%$, implying that they may be analogues to high-redshift quiescent galaxies that formed on similar timescales. Future analysis of this unprecedented sample promises to illuminate the physical mechanisms that drive the quenching of massive galaxies after cosmic noon.

\end{abstract}

\keywords{Post-starburst galaxies (2176), Galaxy quenching (2040), Galaxy evolution (594), Quenched galaxies (2016), Galaxies (573)}


\section{Introduction} \label{sec:intro}

In the local Universe, the vast majority of massive (log$(M_\star/M_\odot) \gtrsim 11$) galaxies are completely quiescent and have been so for $5-10$ Gyr \citep[e.g.,][]{Muzzin2013, Donnari2019, Leja2021, Weaver2022}. There is a growing consensus that two distinct pathways to quiescence are at play, with a rapid path dominating the buildup of quiescent galaxies at high redshifts and a slower channel that populates the ``green valley" at low redshift \citep[e.g.,][]{Schawinski2014,Wu2018,Maltby2018, Belli2019, Suess2021}. While the observational evidence for more rapid early star-formation in the most massive systems at early times is strong \citep[e.g., ``downsizing" trends observed in][]{Juneau2005}, the precise details of how the quiescent population grows from the rapid quenching pathway as a function of cosmic time remain very uncertain. Some studies have characterized the rates of rapid quenching as a function of cosmic time using either photometric \citep{Whitaker2012a,Wild2016, Belli2019, Park2022} or shallow spectroscopic \citep{Rowlands2018b} samples and have found that recently quenched galaxies, sometimes known as post-starburst galaxies, stopped contributing significantly to the quiescent population by $z\gtrsim0.5$. However, photometric studies yield weak constraints on timescales and star formation histories. Thus, our picture of precisely when galaxies shut off their star formation and the contribution of late-time star formation remains poorly constrained.

Ideally, one would study the assembly of the red sequence by modeling the star-formation histories of complete samples of massive galaxies and studying how the incidence and characteristics of the population vary as a function of cosmic time. An immense amount of work has been done over the past decades to study that star formation histories of quiescent systems across cosmic time using photometric and spectroscopic data \citep[e.g.,][]{Tinsley1976,Dressler2004,Dressler2016,Gallazzi2005, Gallazzi2014,Daddi2005,Pacifici2016, Carnall2019, Belli2019,Tacchella2022}. However, measuring the high-order moments of a star-formation history, such as timescales and burst fractions, requires high signal-to-noise continuum spectroscopy \citep{Suess2022b}. The limiting factor in performing such modeling has been the availability of sufficiently deep spectra beyond $z\gtrsim0.5$. The largest existing spectroscopic samples have not prioritized observing the gamut of quiescent galaxies; the SDSS LRG \citep{Eisenstein2001} and BOSS \citep{Dawson2013} surveys targeted the reddest quiescent galaxies, prioritizing pure, uniform samples at the expense of younger, bluer galaxies, with targeting that steeply drops off at $z\sim0.5$ where the post-starburst population beings to emerge \citep{Wild2016, Belli2019}. In contrast, the EBOSS \citep{Dawson2016} survey poorly sampled the quiescent population in favor of more accessible emission line sources. Deeper, more targeted surveys such as LEGA-C \citep{VanDerWel2021, Wu2018}, Carnegie-Spitzer-IMACS \citep{Dressler2016}, and VANDELS \citep{McLure2018, Carnall2019} have identified samples of $\sim$1000s of massive quiescent galaxies at $z\gtrsim0.5$, requiring significant investments on deep fields to reveal spectroscopic information for small samples. 

The next generation of large spectroscopic surveys will revolutionize the availability of continuum spectroscopy of massive galaxies. Here, we utilize the Dark Energy Spectroscopic Instrument (DESI), a robotic, fiber-fed, highly multiplexed spectroscopic surveyor that operates on the Mayall 4-meter telescope at Kitt Peak National Observatory \citep{DESI2016a}. DESI, which can obtain simultaneous spectra of almost 5000 objects over a $\sim3 ^{\circ}$ field \citep[][Miller et al. in preparation]{DESI2016b,Silber2022}, is currently over a year into a five-year survey of approximately one-third of the sky \citep{DESI2016a}, and has already observed more galaxies than the entire Sloan Digital Sky Survey. The DESI Luminous Red Galaxy (LRG) target selection is both broader in color and faintness relative to surveys like BOSS, and as a result is complete to higher redshift ($z\sim0.8$) and observes the Balmer break out to $z\sim1.3$ \citep{Zhou2022}. Here we show that even the relatively small ($\sim20000$ galaxies) but deep Survey Validation (SV) sample of LRGs within the DESI Survey can be leveraged to identify new and exciting samples of recently quenched galaxies that push well beyond what previous surveys have been capable of. 


In this letter, we infer non-parametric star formation histories of LRGs in the DESI SV sample (DESI Collaboration et al. in preparation) and use them to study the growth of the red sequence from recently quenched galaxies. In Section \ref{sec:data}, we describe the parent sample and demonstrate the use of non-parametric star formation histories to fit the spectrophotometric data with \texttt{Prospector} \citep{Johnson2017, Leja2017,Johnson2021}. In Section \ref{sec:analysis}, we use the results of this fitting to identify recently quenched galaxies and characterize their evolving number densities as a function of cosmic time. Finally, in Section \ref{sec:discussion}, we discuss the implications of these findings on our understanding of the physical mechanisms that are driving the production of massive, quiescent galaxies through the rapid quenching channel. 

Throughout this letter, we compare our own selection of ``recently quenched galaxies" to literature samples and selection criteria for post-starburst galaxies. We note that many of these post-starburst selections do not explicitly require a burst of star formation, as any dramatic truncation in star formation can produce an A-star dominated SED.. We assume a concordance $\Lambda$CDM cosmology with $\Omega_{\Lambda}=0.7$, $\Omega_m=0.3$ and $H_0=70$ $\mathrm{km\,s^{-1}\,Mpc^{-1}}$, and quote AB magnitudes.

\section{Data} \label{sec:data}

\subsection{The DESI LRG SV Sample}

In order to characterize the growth of the population of quiescent galaxies at intermediate redshifts, this work relies on the large program of deep spectra that were taken as a part of the DESI SV LRG sample \citep[][DESI Collaboration et al. in preparation]{Zhou2020, Zhou2022}. The primary objective of DESI is to determine the nature of dark energy with precise cosmological measurements \citep{Levi2013}, but the wealth of spectroscopy provides an excellent sample for studies of galaxy evolution. The data volume of the DESI requires multiple supporting software pipelines and products used in this work. Target selection and photometry, which included forward modeling of the differential effect of the PSF across bands, was performed on imaging from the DESI Legacy Surveys \citep[Schlegel et al. in preparation]{Zou2017,Dey2019}. Fiber assignments, tiling, and target selection were performed with the algorithms outlined in Raichoor et al., Schlafly et al., and Myers et al. (in preparation) respectively. All redshifts were determined with the \texttt{Redrock} pipeline (Bailey et al. in preparation). All spectroscopy used was reduced using the ``Fuji'' internal spectroscopic data release which will be identical to the DESI Early Data Release \citep[][EDR, tentatively expected in early 2023]{Guy2022}. 

There are two primary reasons for the choice to utilize the SV sample. First, the SV selection is more inclusive than subsequent Survey Validation samples and the main DESI sample \citep[see Appendix A in][]{Zhou2022}. While this was intended as a test of the redshift recovery so that targeting could be refined from the main survey, these expanded color cuts mitigate potential bias against observing young, recently quenched LRGs. Second, the SV observations were an order of magnitude deeper than the observations for the main survey, with $\sim2.5$ hours of integration per spectrum, resulting in the high signal-to-noise measurements of the continuum. While the SV sample included fainter targets, we restrict this study to the brightest SV LRGs with an observed fiber z magnitude $z_\mathrm{fiber}<21.6$ cut similar to the one that is used in the full LRG sample.

We select all tiles that were observed under the dark time observing conditions in SV. We then select all galaxies which meet the LRG SV cuts outlined in \cite{Zhou2022} with an additional $z_\mathrm{fiber}<21.6$ magnitude constraint, a cut at $z>0.4$ (above which the SV LRG sample begins to be mass complete), and a cut at $z<1.3$ (at which point the age-sensitive $H_\delta$ absorption feature is no longer covered by DESI spectroscopy). We remove galaxies with poor redshift measurements by applying a cut of \texttt{ZWARN} $==0$ to the DESI catalog. We then remove the 580/17797 galaxies that did not reach target depth (exposure time $t_\mathrm{exp}>1$ hour). The median exposure time of this final sample is 2.4 hours, with 16th and 84th percentile exposure times of 1.5 and 4.1 hours respectively. This selection results in a total sample of 17217 galaxies. 


\begin{figure*}
    \includegraphics[width=\textwidth]{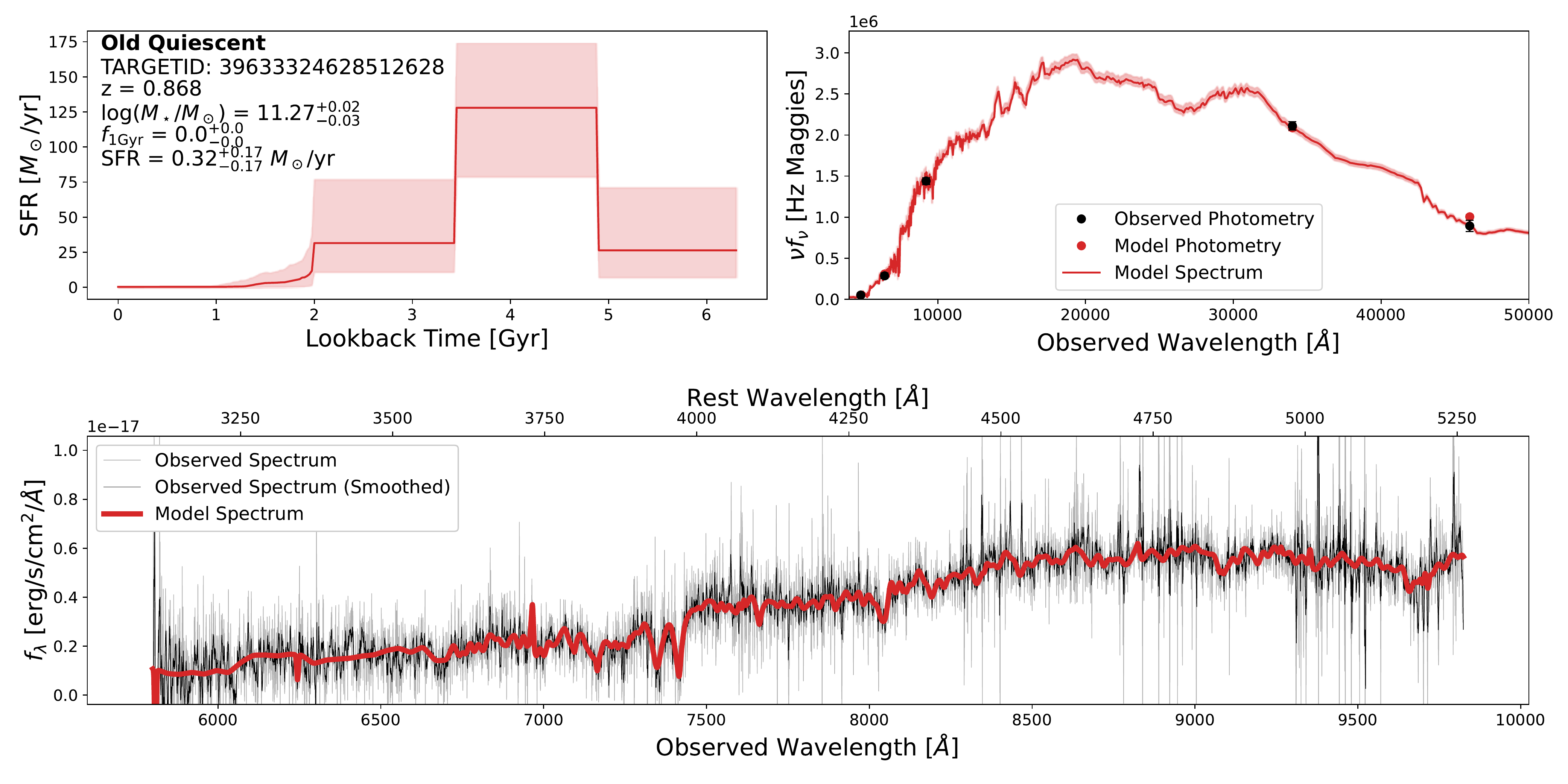}
    \includegraphics[width=\textwidth]{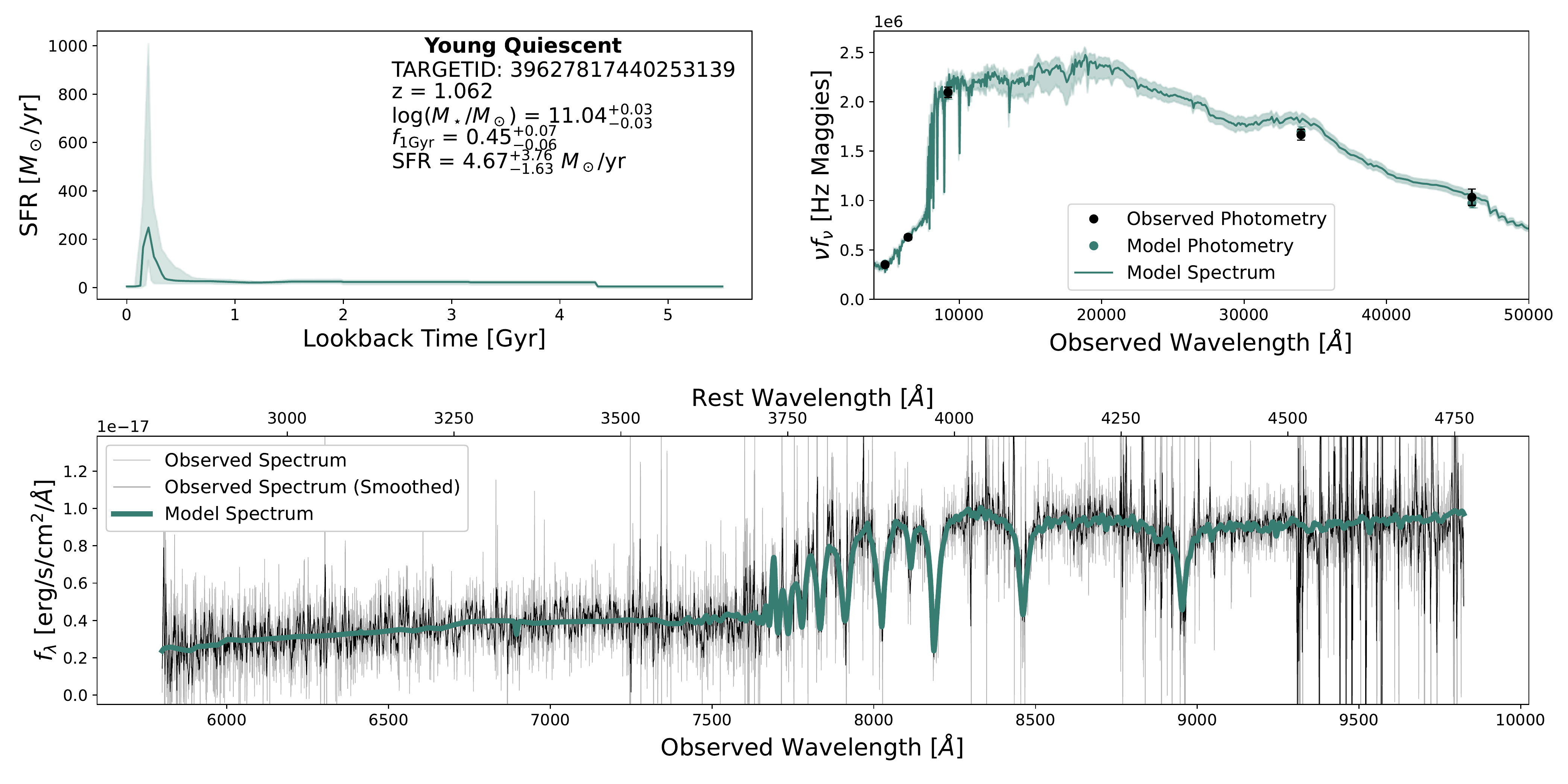}
    \caption{Example old (top, TARGETID=3963332462851262) and recently quenched (bottom, TARGETID=39627817440253139) galaxies from the DESI SV LRG Sample with \texttt{Prospector} fits using the star-formation history model from \cite{Suess2022b}. For each galaxy, we show the median and 68\% confidence interval star-formation history (top left) with selected galaxy properties. We also show the best fitting models (color) to the observed photometry ($g/r/z/W1/W2$, black) (top right). Finally, we show the DESI spectrum (observed, grey; 5 pixel boxcar smoothed, black) along with the best fitting model (color) (bottom). From this modeling, we identify quiescent LRGs and infer the dominance of recent star formation and the timescale of quenching.} 
    \label{fig:prospector_fit}
\end{figure*}

\subsection{Inferring Star Formation Histories with \texttt{Prospector}}

We model the DESI spectra and photometry using non-parametric star formation histories with the SED fitting code \texttt{Prospector} \citep{Johnson2017, Leja2017,Johnson2021} to infer the detailed stellar populations of the sample. Non-parametric star-formation histories (SFHs) are particularly useful for fitting post-starburst galaxies because they do not impose an analytic form on the shape of the SFH, which allows for multiple rises and falls over the course of a galaxy's lifetime. We adopt a flexible bin model that is optimized to model recently quenching galaxies \citep{Suess2022b}. The model utilizes 3 fixed time bins at early times ($t_\mathrm{lookback}>2$ Gyr), 5 flexible bins that each form the same amount of total stellar mass (allowing for greater resolution near periods of intense star-formation), and a final flexible bin that allows a galaxy to remain quenched after star formation is finished. This scheme was extensively tested and is well designed to recover quenching timescales and burst mass fractions \citep{Suess2022b, Suess2022a}.

We use the \texttt{dynesty} dynamic nested sampling package \citep{Speagle2020}, the Flexible Stellar Population Synthesis (FSPS) stellar population synthesis models \citep{Conroy2009, Conroy2010}, the MILES spectral library \citep{Sanchez-Blazquez2006,Falcon-Barroso2011}, and the MIST isochrones \citep{Choi2016, Dotter2016}. We assume a \cite{Chabrier2003} Initial Mass Function and fix the model redshift to the spectroscopic redshift. In contrast with the \cite{Suess2022a} prescription for fitting post-starburst galaxies, we elect to fit nebular emission non-physically by marginalizing over Gaussian lines at the locations of emission features in the spectrum. The massive LRG sample likely hosts many active galactic nuclei (AGN) which can contribute strongly to a galaxy's emission line strength \citep[especially the recently quenched galaxies, e.g.,][]{Greene2020}. Additionally, the LRG selection allows for the targeting of a small fraction of dusty star-forming galaxies with strong emission lines; we want to be completely agnostic to the source of emission when fitting star formation histories to these galaxies. This procedure subtracts out the emission from the spectrum at each step in the fitting before calculating the likelihood, which allows the fits to utilize continuum information (e.g. $H\beta$ absorption) despite the existence of emission that our models do not generate using information about the current SFR. 

We use the mass–metallicity prior described in \cite{Leja2019}. We utilize the \texttt{PolySpecModel} procedure which accounts for deviations between the shape of the photometry and the spectrum by dividing out a polynomial from the observed and model spectra during fitting, using a \texttt{Prospecter}-default 12th order Chebyshev polynomial. We assume the \cite{Kriek2013} dust law with a free $A_v$ and dust index. Additionally, following \cite{Wild2020}, we assume that the attenuation is doubled around young ($<10^7$ yr) stars. We fix the shape of the IR SED following the \cite{Draine2007} dust emission templates, with $U_\mathrm{min}$ = 1.0, $\gamma_e$ = 0.01, and $q_\mathrm{PAH}$ = 2.0. Finally, we include both a spectroscopic jitter term to account for the possibility of underestimated noise and the \texttt{Prospector} pixel outlier model. We center priors on the SFH such that they follow the predicted SFH of a massive quiescent galaxy from the \texttt{UNIVERSEMACHINE} catalog \citep{Behroozi2019}; this weakly prefers solutions with early-time star formation in the star formation histories we fit to ensure that outshining of a young stellar population is treated conservatively. The fidelity of this star formation history at recovering mock parameters is illustrated in \cite{Suess2022b}. Of principle importance to this work, the burst fraction is well recovered when $<50\%$ of a galaxy's stellar mass is formed in a burst. For greater burst fractions, outshining by the young stellar population becomes so dominant that the relative strength of the oldest stellar population cannot be constrained by the existing data, and as such, our conservative prior drives the fits to a higher burst fraction solution than the inputs. Thus, burst fractions measured in this work to be $\gtrsim$50\% can be thought of as strong lower limits.


We fit all \nlrgtot galaxies in the DESI SV LRG sample ($z_\mathrm{fiber}<21.6$) with this procedure, providing the Milky Way extinction corrected $g/r/z/W1/W2$ photometry (using the extinction maps from \citealt{Schlegel1998}) and the galaxy spectrum. The scaling of the SED being fit is set by the photometry that captures all galaxy light rather than just the light in the fiber. We expect the total fraction of galaxy light contained in the fiber to vary as a function of redshift but to always be $\gtrsim 50\%$ of the total light, as the fiber size is 0.75" in radius (~4 kpc at z=0.4, ~6.5 kpc at z=1.3). Our fits are constrained by the SED shape of the photometry , and the polynomial correction to the spectrum will account for any color gradients, though we expect those to be minimal given that both the spectrum should be representative of the majority of the galaxy light for most of the sample. Because the signal in the redshift range of interest is concentrated at the red end of the spectrograph, we elect to only fit the spectra from the R and Z arms of the spectrograph ($5800 \mathrm{\AA}<\lambda_\mathrm{obs} < 9824 \mathrm{\AA}$) to save on computation time and to avoid any issues with the flux calibration at the fainter end of the spectra. In this wavelength range, the resolution R ($\lambda/\Delta\lambda$) ranges from $\sim3200-5100$. While the 1.5'' ($\sim$8 kpc at $z=0.4$, $\sim$13 kpc at $z=1.3$) diameter aperture of the DESI fiber is large enough to capture the majority of galaxy light at highest redshift end of our sample, we do note that our modeling approach assumes a lack of color gradients in the galaxies and that the light represented in the spectrum is identical to that of the photometry, which models all galaxy light. Fits failed to converge for 52/17217 galaxies (0.3\% of the total sample). Visual inspection of the spectra of these failed fits suggests that they broadly fall into four categories: extremely low signal-to-noise galaxies, spectra with large masked regions, galaxies with incorrect redshift assignment, and broad-line AGN/QSOs (which our models are not equipped to characterize). As such, we omit the unmodeled galaxies and perform all analysis on the 17703 successfully fit galaxies. 

\begin{deluxetable}{ccccc}
\tabletypesize{\scriptsize}
\tablecaption{Selected Fit Quantities and Errors\label{tbl:fits}}
\tablehead{\colhead{z} & \colhead{\logM} & \colhead{SFR [$M_\odot$/yr]} & \colhead{\deltasfr} & \colhead{\fonegyr}}
\startdata
0.5568 & 11.23$_{-0.01}^{+0.01}$ & 2.31$_{-0.35}^{+0.47}$ & -1.16$_{-0.07}^{+0.08}$ & 0.11$_{-0.01}^{+0.01}$\\
0.6701 & 11.22$_{-0.01}^{+0.01}$ & 1.17$_{-0.14}^{+0.14}$ & -1.52$_{-0.06}^{+0.05}$ & 0.0$_{-0.0}^{+0.0}$\\
0.8976 & 11.23$_{-0.1}^{+0.02}$ & 0.92$_{-0.31}^{+0.84}$ & -1.76$_{-0.18}^{+0.39}$ & 0.0$_{-0.0}^{+0.07}$\\
0.5396 & 11.36$_{-0.01}^{+0.01}$ & 2.98$_{-0.34}^{+0.4}$ & -1.16$_{-0.06}^{+0.06}$ & 0.05$_{-0.02}^{+0.01}$\\
0.4364 & 11.12$_{-0.01}^{+0.01}$ & 2.85$_{-0.24}^{+0.27}$ & -0.9$_{-0.04}^{+0.05}$ & 0.01$_{-0.0}^{+0.0}$\\
0.8807 & 11.34$_{-0.03}^{+0.03}$ & 0.08$_{-0.08}^{+0.28}$ & -2.86$_{-1.51}^{+0.62}$ & 0.0$_{-0.0}^{+0.0}$\\
0.6999 & 10.8$_{-0.04}^{+0.03}$ & 19.39$_{-2.4}^{+2.56}$ & 0.08$_{-0.09}^{+0.07}$ & 0.22$_{-0.04}^{+0.05}$\\
0.5415 & 11.11$_{-0.03}^{+0.04}$ & 0.09$_{-0.04}^{+0.05}$ & -2.45$_{-0.28}^{+0.19}$ & 0.01$_{-0.0}^{+0.01}$\\
0.5166 & 11.2$_{-0.03}^{+0.02}$ & 21.25$_{-3.02}^{+2.17}$ & -0.15$_{-0.08}^{+0.06}$ & 0.1$_{-0.02}^{+0.02}$\\
1.0623 & 11.04$_{-0.03}^{+0.03}$ & 4.67$_{-1.63}^{+3.76}$ & -0.94$_{-0.18}^{+0.26}$ & 0.45$_{-0.06}^{+0.07}$
\enddata
\tablecomments{Selected median and 68\% confidence values of relevant parameters derived from the posteriors of the \texttt{Prospector} fits to DESI SV LRGs for a random sample of galaxies.}
\vspace{-20pt}
\end{deluxetable}

Example fits to quiescent (top, red) and recently quenched (bottom, green) galaxies are shown in Figure \ref{fig:prospector_fit}. The quiescent galaxy that is representative of the majority of the DESI LRG sample is fit entirely with early star formation, consistent with a very old stellar population, and as such, all the mass was formed in the three fixed-width early-time bins. In contrast, the recently quenched galaxy is clearly fit with a post-starburst SED shape with strong Balmer absorption features and a characteristic lack of emission line infill. This indicates that the post-starburst galaxy has quenched after a period of intense star formation, and the star formation history reflects this. We infer that the galaxy began rapidly forming stars $\sim500$ Myr before observation, and quenched $\sim150$ Myr ago.

From the posteriors on the star formation histories, we derive a number of model parameters, many of which we directly use to select and characterize the properties of recently quenched galaxies. Stellar masses are calculated accounting for mass loss and have typical $1\sigma$ uncertainties of 0.025 dex, and rest absolute magnitudes are calculated directly from the spectra generated from the posterior. We measure the star-formation rate in all galaxies as the star-formation rate in the closest bin to the epoch of observation in the non-parametric star formation history. Above $\sim1 M_\odot$/yr, these star-formation rates have been shown to reliably recover the instantaneous star-formation rate of mock galaxies, and our measurements have typical uncertainties of $\sim15\%$. Below this, they are effectively upper limits \citep{Suess2022b}. Additionally, we quantify the offset from the star-forming sequence, \deltasfrcomma, as:

\begin{figure*}[ht!]
    \includegraphics[width=\textwidth]{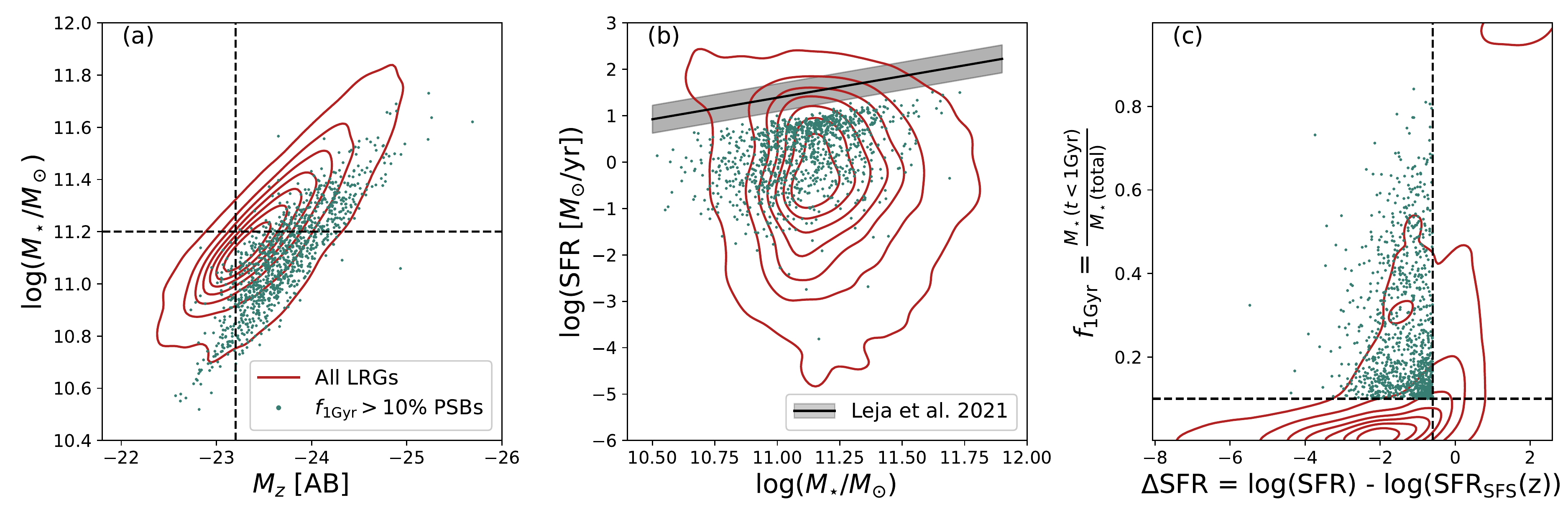}
    \caption{Properties of the full LRG sample (red contours) and a subset of galaxies that recently quenched a significant episode of star formation using our fiducial selection ($f_\mathrm{1 Gyr}>0.1$, $\Delta$SFR$<-0.6$, green points). All plotted points are the median values from the posterior of the \texttt{Prospector} fits. In panel (a), we show the stellar mass versus the absolute magnitude ($M_z$) along with the magnitude limited threshold ($M_z<-23.2$) and the mass complete threshold (log($M_\star/M_\odot > 11.2$) discussed in Section \ref{subsec:subsamples}. In panel (b), we show the star-formation rate versus stellar mass, with the star-forming sequence at $z=0.7$, the median redshift of our sample, shown as a black line with characteristic $\sim0.3$ dex $1\sigma$ scatter \citep{Leja2021}. In panel (c), we show the recently quenched selection plane, \fonegyr versus \deltasfrcomma, with the fiducial selection cuts illustrated as dashed lines. The  sample is significantly brighter than the parent sample at fixed stellar mass and occupies a unique part of parameter space by having formed a significant amount of recent stellar mass despite being fully quenched.}
    \label{fig:PSB_selection}
\end{figure*}

\begin{equation}
    \mathrm{\Delta SFR = log(SFR) - log(SFR_{SFS}(z)})
\end{equation} 

\noindent where $\mathrm{SFR_{SFS}(z)}$ is the inferred star formation rate from the star-forming sequence at the observed redshift of the galaxy defined in \cite{Leja2021}, which is also measured using \texttt{Prospector} SED fits. We set a fiducial threshold for quiescence at $\Delta \mathrm{SFR}=-0.6$, $\sim 2 \sigma$ below the main sequence at a given redshift. Near the fiducial value, the typical uncertainty in \deltasfr is $\sim 0.1$ dex. As with the star formation rate, this value is significantly more uncertain for measured values. Finally, we measure the fraction of the total stellar mass formed in the Gyr before observation, $f_\mathrm{1 Gyr}$. Galaxies with small \fonegyr are very well constrained to be small, and for galaxies which formed $10-70\%$ of their stellar mass in the past Gyr, typical uncertainties are $15-30\%$. A sample of constraints on parameters is shown in Table \ref{tbl:fits}.

We show some of the observed and derived characteristics of the full LRG sample as red contours in Figure \ref{fig:PSB_selection}. In the first panel, we show the stellar mass versus the rest frame absolute magnitude, $M_z$, illustrating the tight correlation between the two parameters. We additionally show lines which correspond to the cuts we make in the two parameters to construct the volume limited samples described in Section \ref{subsec:subsamples}. In the next panel, we show the star formation rate versus the stellar mass along with the ``star forming sequence" at $z=0.7$ with 0.3 dex scatter from \cite{Leja2021} to illustrate that the sample is largely quiescent. Finally, we show the sample in the selection plane of \fonegyr versus \deltasfr discussed in Section \ref{subsec:selection} with our fiducial cuts to select recently quenched galaxies. In all 3 planes, we show the fiducial sample of  galaxies as green points.

\subsection{Selecting Volume Limited Samples} \label{subsec:subsamples}

Because the choices made in spectroscopic targeting significantly impact the observed sample, it is necessary to select a volume limited sample to fairly compare galaxies across redshift bins. This is especially true because the $z_\mathrm{fiber}<21.6$ cut in observed magnitude would observe a faint galaxy at low-redshift but not high-redshift. We use the fits to the spectrophometric data to select samples which we can use to infer number densities. Throughout this letter, we utilize three relevant samples: the full LRG sample, the rest absolute Z-magnitude selected ``magnitude limited" sample, and the ``mass complete" sample to select recently quenched galaxies.

\subsubsection{The Magnitude Limited Sample} \label{subsubsec:mag}

By virtue of being the youngest and brightest galaxies in any given quiescent sample,  galaxies have the lowest $M_\star/L$ ratios at fixed stellar mass and therefore are relatively bright compared to the majority of LRGs. As such, in order to get large, complete samples of  galaxies to study as a function of redshift, a luminosity cut will maximize the sample size. We define a magnitude limited sample with rest-frame $M_z<-23.2$, at which the entirety of the reddest (in rest $g-z$ color, which should map to the highest $M_\star/L$ ratios) 2.5\% of the LRG sample is selected at $z=0.8$. This selection results in the largest volume limited sample we can obtain where we expect to have observed all bright  galaxies in DESI out to $z\sim0.8$, yielding a total of 8683 galaxies.

\begin{figure*}
    \includegraphics[width=\textwidth]{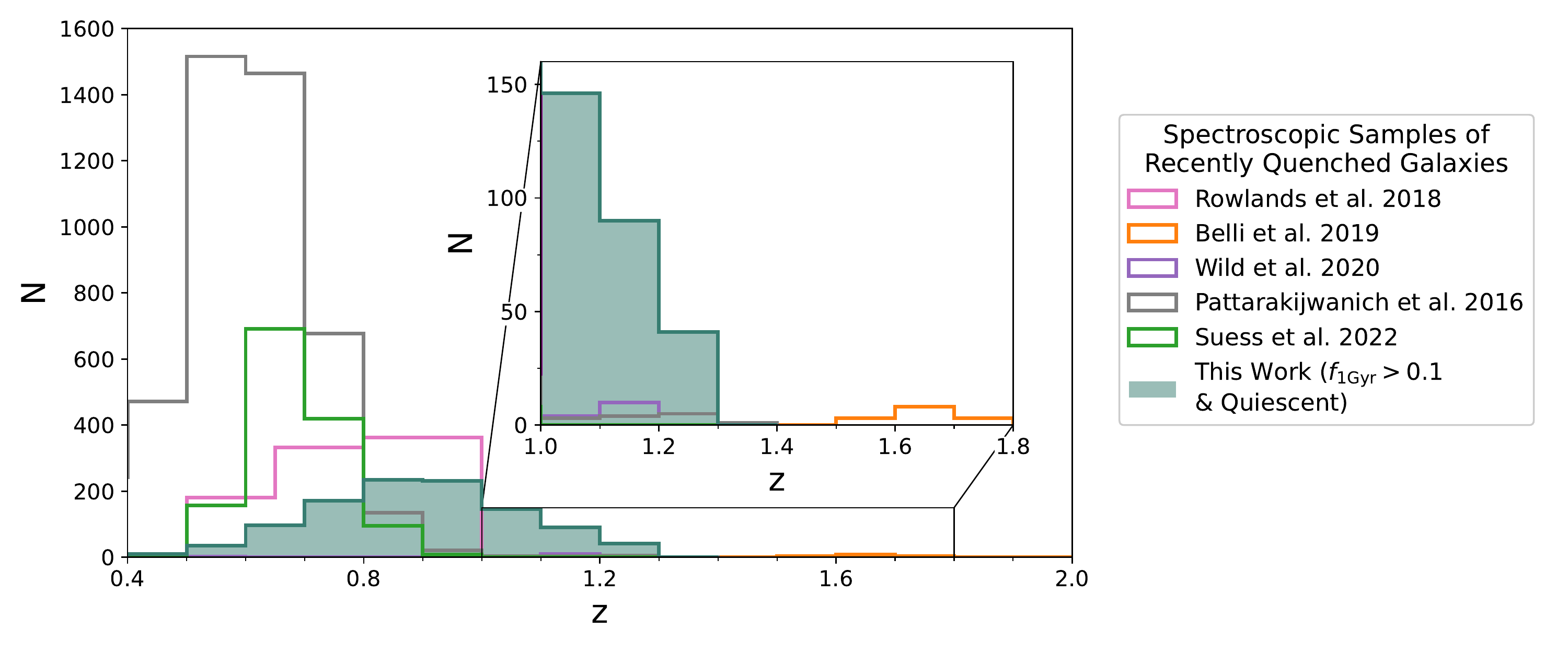}
    \caption{Redshift distributions of spectroscopic samples of recently quenched galaxies from $0.4<z<2.0$, with an inset focusing on $z>1$ where the improvement in sample size from this work is most significant. Our fiducial  sample (\fonegyr$>0.1$, \deltasfr$<-0.6$, selected from the full LRG sample) is shown as a filled green histogram. Other samples shown include PCA-identified post-starburst galaxies from \cite{Rowlands2018b} and \cite{Wild2020}, galaxies with $t_{50}<1.5$ Gyr from \cite{Belli2019}, galaxies selected with K+A template fitting from the SDSS \cite{Pattarakijwanich2016}, and galaxies selected using rest \textit{UBV} filters from the \squiggle sample also selected from the SDSS \citep{Suess2022a}.}
    \label{fig:redshift_distributions}
\end{figure*}

\subsubsection{The Mass Complete Sample} \label{subsubsec:mass}

While a magnitude limited sample selects the bulk of the  galaxies in the SV sample, in order to characterize the growth of the  population relative to the fainter (at fixed stellar mass) old quiescent population, we instead require a mass complete sample. In the redshift range $0.4<z<0.8$, the DESI LRG targeting only selects a sample which is $\gtrsim 80\%$ mass complete for very massive galaxies \citep[\logM$\gtrsim11.2$--accounting for systematic differences between the stellar masses we measure and those in][]{Zhou2022}. As such, in situations where we wish to compare to the quiescent population as a whole, we elect to use only galaxies above this stellar mass, regardless of their rest-frame $M_z$. This sample is significantly smaller than the magnitude limited sample, with only 5375 galaxies above the stellar mass cut at $z<0.8$.

\vspace{8pt}

We show the cuts in rest-frame $M_z$ and stellar mass that result in the two subsamples in Figure \ref{fig:PSB_selection}a, illustrating that the stellar mass cut is significantly more restrictive than the magnitude cut, which lets through  galaxies at masses as low as $10^{10.8} M_\odot$. At fixed stellar mass, the fiducial  sample (see Section \ref{subsec:selection}) is significantly brighter than a typical LRG (red contours), and we therefore maximize our ability to constrain the number density of recently quenched galaxies as a population by instituting a cut on the absolute magnitude.

\section{Analysis} \label{sec:analysis}

\subsection{Selecting Recently Quenched Galaxies \label{subsec:selection}}

There are a number of ways of selecting recently quenched/post-starburst galaxies, all of which share the common goal of selecting galaxies that recently quenched after a period of significant star formation \citep{French2021}. Historically, these galaxies have been selected using a combination of emission line cuts (to select against current star formation) and Balmer absorption depth (to select for a stellar population dominated by A type stars) \citep{Dressler1983,Zabludoff1996,Balogh1999}. Here, we leverage the tightly constrained star formation histories to select a physically motivated sample of recently quenched galaxies. First, we focus on selecting a pure quiescent sample. In Figure \ref{fig:PSB_selection}b+c, it is clear that some galaxies that are dusty and star-forming have been selected due to their red colors and exist in the LRG parent sample. To remove these, we perform a conservative cut in \deltasfrcomma, classifying galaxies as quiescent only if their median \deltasfr is $\sim2\sigma$ (0.6 dex) below the star-forming sequence at their redshift from \cite{Leja2021}. This selection, which is highlighted in Figure \ref{fig:PSB_selection}c, removes 2622 galaxies ($\sim15\%$ of the total sample). All qualitative results in this work are insensitive to the exact definition of quiescence that we adopt, though exact sample sizes and number densities will by definition differ slightly.

Secondly, we are interested in separating the quiescent galaxy population physically into recent additions to the red sequence and older galaxies. In this work, our definition of recently quenched does not require a burst, as we are interested in classifying all galaxies which rapidly formed a significant amount of stellar mass before quenching as  galaxies. To select such a sample, we leverage the inferred star formation histories to measure the fraction of the stellar mass formed within the last Gyr (\fonegyrcomma) for all galaxies \citep[see also][]{Webb2020}. In combination with the cut for quiescence, selecting galaxies with high \fonegyr identifies a sample that must have rapidly truncated its star formation in order to have formed a large amount of its stellar mass while also reaching quiescence within 1 Gyr. We adopt \fonegyr$>0.1$ (also shown in Figure \ref{fig:PSB_selection}c) for our fiducial  selection and explore the impact of different thresholds in Section \ref{sec:discussion}. The fiducial selection identifies 1089  galaxies from the 15012 quiescent LRGs using the fiducial \fonegyr$>0.1$ selection. 

\begin{figure*}
    \includegraphics[width=\textwidth]{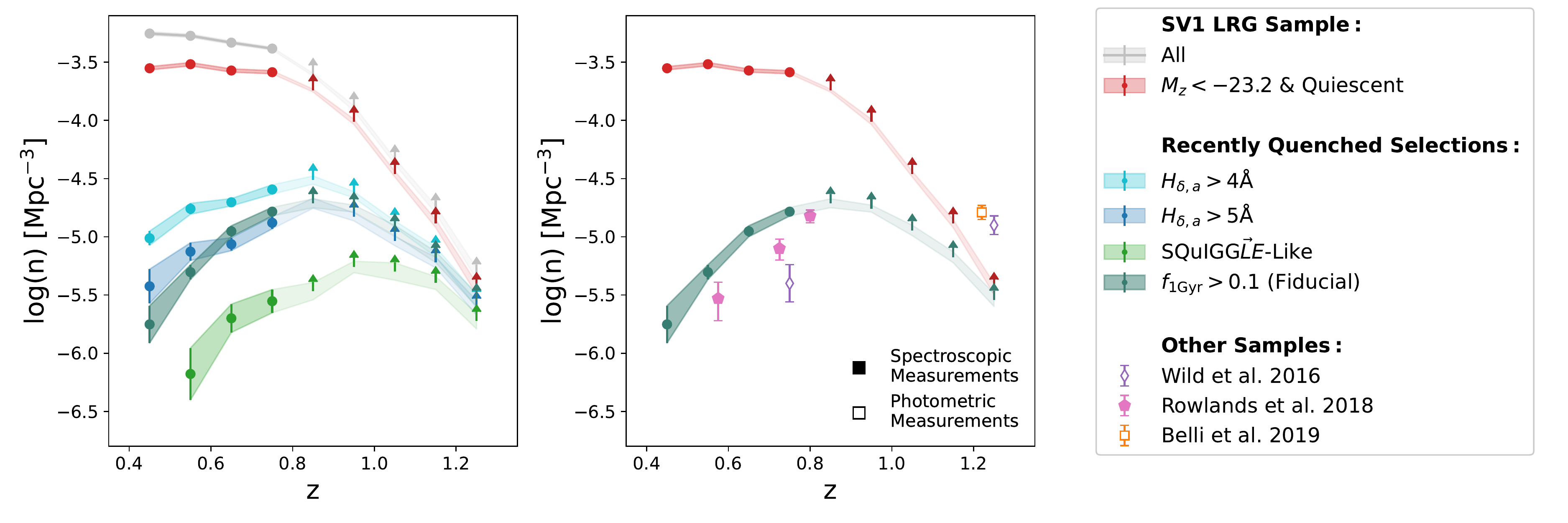}
    \caption{(Left): Number densities within the DESI SV LRG sample (full sample, gray; luminosity-complete and quiescent, red) and a variety of  selections from the magnitude limited (see Section \ref{subsubsec:mag}) quiescent sample ($H_{\delta,a}>4$, light blue; $H_{\delta,a}>5$, dark blue; SQuIGG$\vec{L}$E SED selection, light green; and $f_\mathrm{1 Gyr}$, green). Beyond $z\sim0.8$, we indicate that the measured number densities are lower limits by plotting as upward facing arrows. All  selections show an increasing number density over the redshift range in which we are complete, with varying normalization resulting from the relative restrictiveness of the post-starburst criteria. (Right): The same magnitude limited LRG and $f_\mathrm{1 Gyr}>0.1$ samples as the previous panel in addition to literature measurements (photometric: open symbols; spectroscopic: filled symbols). All three of the \cite{Wild2016} ($M_\star>10^{10.8} M_\odot$), \cite{Rowlands2018b} ($M_\star>10^{11} M_\odot$), and \cite{Belli2019} ($M_\star>10^{10.8} M_\odot$) samples show a trend of increasing number density with redshift, but the normalization differs between the different samples as a result of differing stellar mass limits and selection techniques.}
    \label{fig:number_density}
\end{figure*}

This sample of  galaxies is unparelled in size beyond $z\gtrsim1$. In Figure \ref{fig:redshift_distributions}, we show the redshift distributions of this sample compared to other large spectroscopic samples of post-starburst galaxies at intermediate redshift. Our sample of 100s of  galaxies at $z<0.8$ is smaller than other samples which select galaxies from the full Sloan Digital Sky Survey \citep{Pattarakijwanich2016,Suess2022a} or VIPERS Survey \citep{Rowlands2018b}. However, at $z>1$ (shown in the inset), we find that this sample dramatically increases the number of spectroscopically confirmed  galaxies at the tail end of cosmic noon.

While our selection of  galaxies relies on our inferred star formation histories, there are many other selections that use empirical measures of spectroscopic features to select post-starburst galaxies \citep{French2021}. We choose a few common post-starburst identification methods and compare the resulting number densities with our fiducial model (see Section \ref{subsec:number_density}). For all literature comparisons, we use the same \deltasfr$\leq -0.6$ criterion for quiescence rather than relying on common empirical metrics like EW $H_\alpha$, which falls out of our spectral window for most of the sample, or EW [OII], which is an uncertain tracer of SFR due to potential contribution from AGN/LINERs. We note that while exact definitions of $H_\delta$ spectral indices vary in the literature (e.g., \cite{Alatalo2016b} uses $H_\delta$, \cite{French2015} uses \hdeltacomma, and \cite{Baron2022a} $H_{\delta,F}$), these differences are subtle. We adopt \hdelta as our preferred definition, as it is optimized for features from A-type stars \citep{Worthey1997}. The three selections we compare to our fiducial selection (\fonegyr$>0.1$ and \deltasfr$<-0.6$) are as follows (all numbers quoted are the raw number of galaxies in the full sample, not in a volume limited sample):

\begin{enumerate}
    \item \hdelta$>4$ \angstromcomma: After applying the quiescence criteria, we select \nhdeltafour galaxies with \hdelta $>$ 4\angstrom following e.g., \cite{French2015, French2018a, Wu2018, Yesuf2022}.
    
    \item \hdelta$>5$ \angstromcomma: We impose a more stringent cut, \hdelta$>$5 \angstromcomma, following e.g., \cite{Alatalo2016b, Baron2022a}, selecting \nhdeltafive post-starburst galaxies.
    
    \item \squiggle Selection: Finally, after applying the quiescence criteria, we use medium band synthetic rest-frame $UBV$ filters to identify post-starburst galaxies with $U-B > 0.975$ and $-0.25 < B-V < 0.45$ following the procedure for selecting galaxies with SEDs dominated by A-type stellar populations \citep{Suess2022a}. We apply these cuts to the median best-fit models because the spectral coverage is not red enough to consistently overlap with the synthetic $V$ filter. This selection finds only \nsquiggle post-starburst galaxies.
    
\end{enumerate}

\subsection{The number density of recently quenched galaxies \label{subsec:number_density}}

The DESI SV LRG selection is designed to have a uniform comoving number density of galaxies at $0.4<z<0.8$, which enables robust determination of number densities of subsets of the spectroscopic sample \citep{Zhou2022}. For this selection, we use the target density of 1439 $\mathrm{deg^{-2}}$ to calculate the number density in bins of $\Delta z=0.1$ in redshift from $z=0.4$ to $z=1.3$ by measuring the density of targets for a given selection criterion and dividing by the volume of the bin. We measure the number densities only for the magnitude limited or mass complete samples. We utilize jackknife resampling of the 31 SV pointings to calculate the errors on the measured number densities. The errors do not account for catastrophic redshift errors, but those should be very rare ($\leq0.5\%$, see \citealt{Zhou2022}) and subdominant relative to cosmic variance and Poisson errors.

\begin{figure*}
    \includegraphics[width=\textwidth]{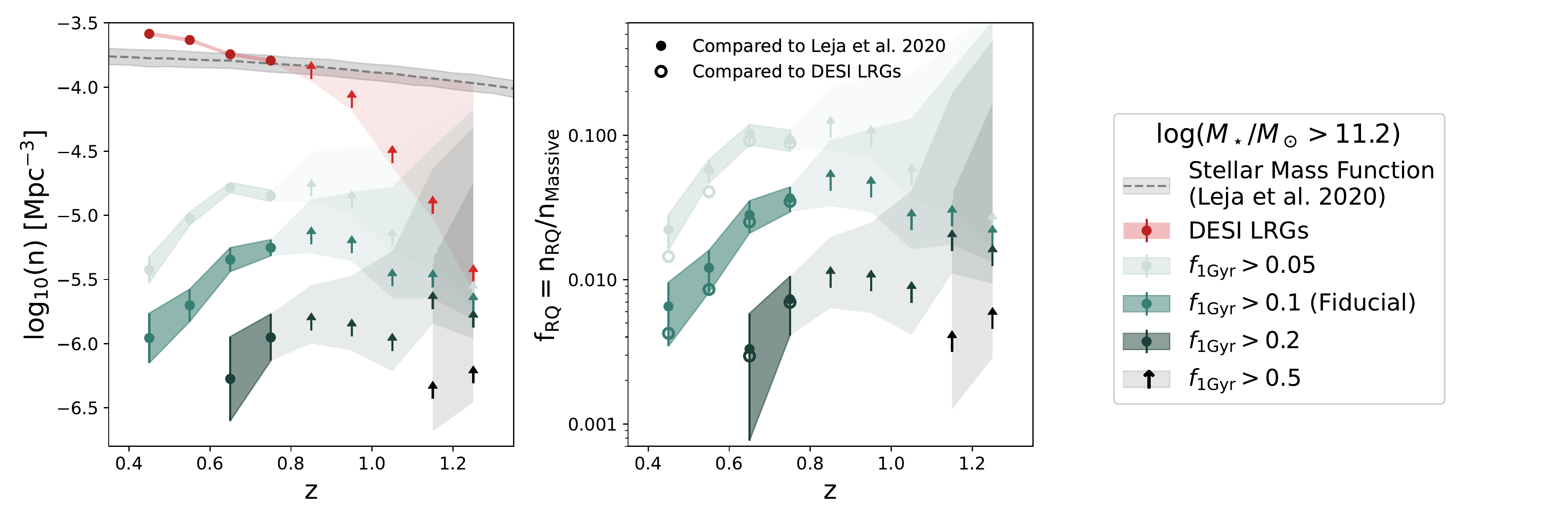}
    \caption{The number densities (left) and fractions (right) of recently quenched galaxies in the mass complete sample (\logM$> 11.2$, see Section \ref{subsubsec:mass}). The dashed line and grey band (left) represent the stellar mass function of similarly mass galaxies \citep{Leja2020}, and the red points show the number density of all LRGs above the mass limit. The light green, green, dark green, and black points represent the number densities and fractions of recently quenched galaxies with \fonegyr greater than 0.05, 0.1, 0.2, and 0.5 respectively, as compared to the stellar mass function from \cite{Leja2020}. On the right panel, the open points of the same colors show the fraction of  galaxies as compared to our own massive galaxy number density measurements. Above $z=0.8$, measurements are indicated as lower limits with errors inflated to encapsulate the possibility that all galaxies which were not targeted by DESI meet the selection criterion.}
    \label{fig:number_density_and_fraction}
\end{figure*}

The comoving number density of each recently quenched subsample of the magnitude limited sample as a function of redshift is shown in Figure \ref{fig:number_density}. The raw number density of the DESI LRG SV sample ($z_\mathrm{fiber}<21.6$) is shown in grey. We show the number density of the rest-frame magnitude limited ($M_z<-23.2$) sample with the fiducial quiescence cut (\deltasfr$<-0.6$) in red. We then apply the post-starburst selections outlined in Section \ref{subsec:selection} to the magnitude limited sample. The number densities are shown for \hdelta $>4$ \angstrom (light blue), for \hdelta $>5$ \angstrom (dark blue),  \squigglecomma-like (light green), and our fiducial $f_\mathrm{1 Gyr}>0.1$ selection (dark green). In all cases, the number density of post-starburst galaxies rises as a function of redshift in the range of redshifts where the parent LRG sample is complete ($z<0.8$). Above this, we illustrate that our measurements are lower limits.

In the second panel of Figure \ref{fig:number_density}, we compare our fiducial sample of  galaxies to several measurements from the literature. We find qualitative agreement with previous studies that observe the number density of recently quenched galaxies increasing with redshift \citep{Wild2016, Rowlands2018b,Belli2019}. Additionally, the number density of the  galaxies that we measure is very similar to that of compact star forming galaxies at z=0.5, adding credence to the argument that such galaxies may be progenitors to local post-starburst galaxies \citep{Tremonti2007, Diamond-Stanic2021, Whalen2022}. However, in detail, this comparison is limited by systematic effects; our sample is systematically more massive than other post-starburst samples, and is selected using a magnitude (not mass) limit. Additionally, as shown in the first panel of Figure \ref{fig:number_density}, differing identification techniques can significantly impact the measured number density of post-starburst galaxies. Still, a clear consensus emerges from this comparison that recently quenched galaxies were increasingly common at greater lookback time.

\subsection{Exploring the Growth of the Red Sequence by Rapidly Quenched Galaxies}

In the previous section, we studied the number density of a magnitude limited sample of  galaxies to maximize our sample size. Here, we attempt to explicitly quantify the fraction of massive galaxies that have recently quenched and joined the red sequence as a function of cosmic time. To do so, we utilize the mass complete (\logM$>11.2$, see Section \ref{subsubsec:mass}) subset of the LRG sample, which we show in the first panel of Figure \ref{fig:number_density_and_fraction} (red) along with the corresponding stellar mass function from \cite{Leja2020}. This measurement over-predicts the stellar mass function by $\sim0.2$ dex at $z\sim0.4$ while matching well at $z\sim0.7$. This may be due to systematic differences in the stellar mass estimates (e.g., differences in modeled star formation histories, unmodeled contributions from AGN, or spectrophotometric modeling in our fits versus broadband multiwavelength SEDs), and the mismatch in redshift evolution may be a result of the targeting incompleteness. As such, we adopt the number densities from the stellar mass function as the total abundance of massive (\logM$<11.2$) galaxies and note that the fractions we measure may be systematically lower than reported by $\sim0.2$ dex. Above $z=0.8$ where LRG targeting is known to be incomplete, we inflate the upper error bar on the measured lower limits by assuming that every galaxy we have not targeted meets the selection criteria (quantified by the deviation between the measured number density and the stellar mass function) to capture the possibility that every galaxy we did not measure is a recently quenched galaxy. Since this is unlikely due to the lower $M_\star/L$ ratio of  galaxies, this conservative estimates captures the full range of possibility in the number density of galaxies in a given selection at $z>0.8$.

We show the number densities of four different selections of recently quenched galaxies: \fonegyr$>0.05$ (light green), \fonegyr$>0.1$ (green), \fonegyr$>0.2$ (dark green), and \fonegyr$>0.5$ (black). Points that do not appear indicate that the redshift bin contained zero galaxies that met the selection criteria. All four sets of recently quenched galaxies show increasing number densities with redshift. However, even at $z\sim0.8$, galaxies which formed a large fraction of their stellar mass in the past Gyr are very rare. For example, at $z=0.8$, galaxies that formed $>20\%$ of their stellar mass in the past Gyr were significantly ($>1$ dex) rarer than those which formed 5\% of their stellar mass. We find that the number density of the $f_\mathrm{1 Gyr}>20\%$ population cannot be decreasing with lookback time, and in fact, at $z\sim1.2$ the lower limit number density of this population is higher than the number density at z=0.8. The rarity of such objects at intermediate-z is extremely consistent with the rarity of ``late bloomers," galaxies that formed the majority of their stellar mass in the 2 Gyr before quenching \citep{Dressler2018}. Additionally, we identify a very small population of galaxies which rapidly formed $\geq50\%$ of their stellar mass in the Gyr before observation at $z=1.1-1.3$, with lower limits that indicate a number density of at least $\mathrm{log_{10}(n)>}-6.5 \ \mathrm{Mpc^{-3}}$. Similar extreme post-starburst galaxies have been found in photometric samples with comparably low number densities and could represent analogs to the formation of massive quiescent galaxies at high-z \citep{Park2022}. 

In the right panel of Figure \ref{fig:number_density_and_fraction}, we show the same  samples as fractions of the total massive galaxy population (shown with solid points using the stellar mass function from \cite{Leja2020} as the denominator and empty points using our own measurements of the LRG number density). We find that galaxies which formed $>20\%$ of their stellar mass represent $\sim0.5\%$ of the total galaxy population at z=0.8, but by $z\sim1.2$ must be at least 1\%, with an upper limit that extends to them being $\sim50\%$ of the quiescent population. Similarly, the most extreme burst-dominated systems (\fonegyr$>50\%$) must be at least $\sim0.5\%$ of the total galaxy population at z=1.2, but this fraction could be as high as 20\%. In contrast, galaxies with \fonegyr$>5\%$ and $>10\%$ are significant even at z=0.4, representing $\sim1.5\%$ and $0.5\%$ of the massive galaxy population, and by $z=0.8$ they are $\sim10\%$ and $3\%$ of the total population. Studies of massive quiescent and post-starburst galaxies at similar redshifts have measured similar burst fractions of $\sim5\%$ in the bulk of their samples, indicating that at the massive end, the vast majority of ``post-starburst" galaxies are the evolutionary products of a recent dusting of star formation rather than the truncation of their primary epoch of star formation \citep{Patel2011, French2018a}.

The general rarity of extreme massive post-starburst galaxies in this sample is consistent with findings that the formation redshift of \logM$=11.2$ galaxies is $z_\mathrm{form}\sim2-3$ \citep{Gallazzi2014,Pacifici2016,Fumagalli2016, Carnall2019,Estrada-Carpenter2019, Diaz-Garcia2019, Webb2020, Khullar2022}; at the epochs we are probing, the average massive quiescent galaxy quenched long in the past. However, we find that a significant number of massive galaxies are still quenching with very high \fonegyr well after cosmic noon ($z\sim2$), and expect that with a more complete sample, at higher redshift the population dominated by recent star formation would become the norm.The sharp observed decline in rapid quenching after cosmic noon suggests a fundamental shift in the evolutionary histories of massive galaxies. By combining this preliminary analysis with similar stellar population synthesis modeling of larger, mass-complete samples and ancillary datasets (e.g., by analyzing galaxy structural evolution and morphological transformation), we hope to illuminate the physical mechanism(s) that are responsible for halting star formation and sustaining quiescence of massive galaxies since $z\sim1$.

\section{Discussion and Conclusions} \label{sec:discussion}

Using the DESI SV sample, we measure non-parametric star formation histories for a novel sample of Luminous Red Galaxies. We select physically motivated samples of  galaxies and leverage the well characterized parent sample to characterize the increasing number density of recently quenched galaxies with lookback time. We find the following:

\begin{enumerate}
    \item The sample of quiescent galaxies which formed $>10\%$ of its stellar mass in the past Gyr represents a novel spectroscopic sample. The sample of 277 galaxies we identify at $z>1$ is an order of magnitude larger than previous samples (see Figure \ref{fig:redshift_distributions}).

    \item The number density of  galaxies rises steadily with redshift from $z=0.4-0.8$ based on our model selection and empirical identification methods; post-starburst galaxies were more common at earlier cosmic time (see Figure \ref{fig:number_density}).
    
    \item The fraction of massive (\logM$>11.2$) galaxies which have recently quenched their star formation and which formed $>10\%$ of their stellar mass in the past Gyr rises in this redshift range from $\lesssim 0.5\%$ at z=0.4 to $\sim3\%$ in at z=0.8 (see Figure \ref{fig:number_density_and_fraction}). Furthermore, at $z>1$, we find a significant emerging population that formed $>20\%$ and $>50\%$ of its stellar mass in the past Gyr. 
    
\end{enumerate}


As our criteria for selecting recently quenched galaxies simply required a rapid truncation in a galaxy's star formation rate that drove a galaxy into quiescence in $<\lesssim$1 Gyr, there is substantial variety in the star formation histories of galaxies which fall into a given selection for \fonegyrcomma. The simplicity of this selection allows for simple determination of the rate at which galaxies have entered into quiescence, but it does not distinguish between, for example, a secondary starburst in an already quiescent galaxy versus a rapid truncation of the primary epoch of star formation at fixed \fonegyrcomma. Future work will endeavor to combine these star formation histories with ancillary data to paint a holistic picture of the quenching of these galaxies. For now, we use constraints on the fraction of galaxies which recently entered into quiescence to discuss possible physical mechanisms that could be driving the rapid cessation of star formation in this sample of galaxies.

One of the most compelling fast processes that could induce, then shut off, star formation and produce post-starburst galaxies is major mergers \citep[e.g.,][]{Hopkins2008}. After cosmic noon, simulations have found that many massive galaxies that quench do so via major mergers \citep[e.g.,][]{Wellons2015}, which funnel gas inward and induce a burst of star formation that rapidly shuts off. Indeed, many studies of post-starburst galaxies have found that merger features are more common in post-starburst systems \citep[e.g.,][]{Pawlik2016,Sazonova2021,Ellison2022,Verrico2022}. 

We estimate the relative frequency of major mergers using the \texttt{UNIVERSEMACHINE} \citep{Behroozi2019} and find that 15\% and 20\% of \logM$>11.2$ galaxies, at $z=0.4$ and $z=0.8$ respectively, experienced a major merger ($M_{\star,2}/M_{\star,1}>25\%$ in the progenitor galaxies in the merger tree) in the past Gyr. This rate is significantly higher than the the 0.5\% and 3\% fractions we find for our fiducial sample of recently quenched galaxies, and the merger fraction increases more slowly than the  fraction. Some of this difference may be driven by gas-poor major mergers between already quiescent systems or gas rich mergers that do not quench, and we conclude that it is plausible that every very massive galaxy that rapidly quenches between $0.4<z<0.8$ does so as a result of a major merger, and that not every major merger results in a post-starburst galaxy. This is in line with predictions from the Illustris TNG simulation that only $\sim5\%$ of massive galaxies will quench within $\sim$500 Myr of coalescence after a major merger \citep{Quai2021}, and indicates that even if major mergers are an essential part of the quenching process, they do not universally produce post-starburst galaxies. However, at high-z, our measured lower limits fall short of placing strong constraints.

Still, the high-z tail of our distribution promises to be a very powerful tool for studying rapid quenching. Prior to DESI, only a small number of spectroscopic continuum observations from surveys could be mined for post-starburst galaxies above $z\gtrsim1$ \citep{Wild2020}, and often samples can only be obtained through targeted followup of photometrically identified sources \citep[e.g.,][]{Belli2019}. Even in the smallest (but highest signal-to-noise) subset of DESI LRG spectra, we have identified an order of magnitude more spectroscopically confirmed  galaxies than had been measured previously. Future work will leverage these star formation histories further to study trends using parameters such as the time since quenching \citep{Suess2022a}, which has been used in post-starburst populations to constrain the evolution of AGN incidence \citep{Greene2020}, sizes \citep{Setton2022a}, molecular gas contents \citep{Bezanson2022, Spilker2022}, and merger fractions \citep{Verrico2022}. Using the combination of the unique spectroscopically derived moments of the star formation history and ancillary data, we hope to place strong constraints on the mechanisms that drive the quenching of massive galaxies as close to cosmic noon as is currently possible. Future surveys, such as PFS \citep{Greene2022} and MOONRISE \citep{Maiolino2020} will extend wavelength coverage into the NIR, pushing farther in redshift to cosmic noon. In conjunction with this sample, comprehensive studies of the properties of  galaxies from $z=0$ to $z=2$ will paint a cohesive picture of the rapid quenching process and its role in producing the present-day quiescent population.

\newpage

\acknowledgements 

This research is supported by the Director, Office of Science, Office of High Energy Physics of the U.S. Department of Energy under Contract No. DE–AC02–05CH11231, and by the National Energy Research Scientific Computing Center, a DOE Office of Science User Facility under the same contract; additional support for DESI is provided by the U.S. National Science Foundation, Division of Astronomical Sciences under Contract No. AST-0950945 to the NSF’s National Optical-Infrared Astronomy Research Laboratory; the Science and Technologies Facilities Council of the United Kingdom; the Gordon and Betty Moore Foundation; the Heising-Simons Foundation; the French Alternative Energies and Atomic Energy Commission (CEA); the National Council of Science and Technology of Mexico (CONACYT); the Ministry of Science and Innovation of Spain (MICINN), and by the DESI Member Institutions: \url{https://www.desi.lbl.gov/collaborating-institutions}.

DS gratefully acknowledges support from NSF-AAG\#1907697 and the PITT PACC graduate fellowship. The results published were also funded by the Polish National Agency for Academic Exchange (Bekker grant BPN/BEK/2021/1/00298/DEC/1), the European Union's  Horizon 2020 research and innovation programme under the Maria Skłodowska-Curie (grant agreement No 754510) and by the Spanish Ministry of Science and Innovation through Juan de la Cierva-formacion program (reference FJC2018-038792-I). DS acknowledges helpful conversations with Joel Leja, Wren Suess, Sirio Belli, and Sandro Tacchella that improved this work. We also thank the reviewer for their deep and thorough reading of the paper, resulting in a significantly improved final product.

The authors are honored to be permitted to conduct scientific research on Iolkam Du’ag (Kitt Peak), a mountain with particular significance to the Tohono O’odham Nation. We direct readers to Astrobite's coverage of the Tohono O’odham People and Kitt Peak National Observatory, as part of their series on the interactions between predominantly western astrophysics observatories and Indigenous communities: \url{https://astrobites.org/2019/08/16/a-tale-of-two-observatories}.

\newpage


\bibliography{all}

\begin{thebibliography}{}
\expandafter\ifx\csname natexlab\endcsname\relax\def\natexlab#1{#1}\fi
\providecommand{\url}[1]{\href{#1}{#1}}
\providecommand{\dodoi}[1]{doi:~\href{http://doi.org/#1}{\nolinkurl{#1}}}
\providecommand{\doeprint}[1]{\href{http://ascl.net/#1}{\nolinkurl{http://ascl.net/#1}}}
\providecommand{\doarXiv}[1]{\href{https://arxiv.org/abs/#1}{\nolinkurl{https://arxiv.org/abs/#1}}}

\bibitem[{{Alatalo} {et~al.}(2016){Alatalo}, {Lisenfeld}, {Lanz}, {Appleton},
  {Ardila}, {Cales}, {Kewley}, {Lacy}, {Medling}, {Nyland }, {Rich}, \&
  {Urry}}]{Alatalo2016b}
{Alatalo}, K., {Lisenfeld}, U., {Lanz}, L., {et~al.} 2016, \apj, 827, 106,
  \dodoi{10.3847/0004-637X/827/2/106}

\bibitem[{{Balogh} {et~al.}(1999){Balogh}, {Morris}, {Yee}, {Carlberg}, \&
  {Ellingson}}]{Balogh1999}
{Balogh}, M.~L., {Morris}, S.~L., {Yee}, H.~K.~C., {Carlberg}, R.~G., \&
  {Ellingson}, E. 1999, \apj, 527, 54, \dodoi{10.1086/308056}

\bibitem[{{Baron} {et~al.}(2022){Baron}, {Netzer}, {Lutz}, {Prochaska}, \&
  {Davies}}]{Baron2022a}
{Baron}, D., {Netzer}, H., {Lutz}, D., {Prochaska}, J.~X., \& {Davies}, R.~I.
  2022, \mnras, 509, 4457, \dodoi{10.1093/mnras/stab3232}

\bibitem[{{Behroozi} {et~al.}(2019){Behroozi}, {Wechsler}, {Hearin}, \&
  {Conroy}}]{Behroozi2019}
{Behroozi}, P., {Wechsler}, R.~H., {Hearin}, A.~P., \& {Conroy}, C. 2019,
  \mnras, 488, 3143, \dodoi{10.1093/mnras/stz1182}

\bibitem[{{Belli} {et~al.}(2019){Belli}, {Newman}, \& {Ellis}}]{Belli2019}
{Belli}, S., {Newman}, A.~B., \& {Ellis}, R.~S. 2019, \apj, 874, 17,
  \dodoi{10.3847/1538-4357/ab07af}

\bibitem[{{Bezanson} {et~al.}(2022){Bezanson}, {Spilker}, {Suess}, {Setton},
  {Feldmann}, {Greene}, {Kriek}, {Narayanan}, \& {Verrico}}]{Bezanson2022}
{Bezanson}, R., {Spilker}, J.~S., {Suess}, K.~A., {et~al.} 2022, \apj, 925,
  153, \dodoi{10.3847/1538-4357/ac3dfa}

\bibitem[{{Carnall} {et~al.}(2019){Carnall}, {McLure}, {Dunlop}, {Cullen},
  {McLeod}, {Wild}, {Johnson}, {Appleby}, {Dav{\'e}}, {Amorin}, {Bolzonella},
  {Castellano}, {Cimatti}, {Cucciati}, {Gargiulo}, {Garilli}, {Marchi},
  {Pentericci}, {Pozzetti}, {Schreiber}, {Talia}, \& {Zamorani}}]{Carnall2019}
{Carnall}, A.~C., {McLure}, R.~J., {Dunlop}, J.~S., {et~al.} 2019, \mnras, 490,
  417, \dodoi{10.1093/mnras/stz2544}

\bibitem[{{Chabrier}(2003)}]{Chabrier2003}
{Chabrier}, G. 2003, \pasp, 115, 763, \dodoi{10.1086/376392}

\bibitem[{{Choi} {et~al.}(2016){Choi}, {Dotter}, {Conroy}, {Cantiello},
  {Paxton}, \& {Johnson}}]{Choi2016}
{Choi}, J., {Dotter}, A., {Conroy}, C., {et~al.} 2016, \apj, 823, 102,
  \dodoi{10.3847/0004-637X/823/2/102}

\bibitem[{{Conroy} \& {Gunn}(2010)}]{Conroy2010}
{Conroy}, C., \& {Gunn}, J.~E. 2010, \apj, 712, 833,
  \dodoi{10.1088/0004-637X/712/2/833}

\bibitem[{{Conroy} {et~al.}(2009){Conroy}, {Gunn}, \& {White}}]{Conroy2009}
{Conroy}, C., {Gunn}, J.~E., \& {White}, M. 2009, \apj, 699, 486,
  \dodoi{10.1088/0004-637X/699/1/486}

\bibitem[{{Daddi} {et~al.}(2005){Daddi}, {Renzini}, {Pirzkal}, {Cimatti},
  {Malhotra}, {Stiavelli}, {Xu}, {Pasquali}, {Rhoads}, {Brusa}, {di Serego
  Alighieri}, {Ferguson}, {Koekemoer}, {Moustakas}, {Panagia}, \&
  {Windhorst}}]{Daddi2005}
{Daddi}, E., {Renzini}, A., {Pirzkal}, N., {et~al.} 2005, \apj, 626, 680,
  \dodoi{10.1086/430104}

\bibitem[{{Dawson} {et~al.}(2013){Dawson}, {Schlegel}, {Ahn}, {Anderson},
  {Aubourg}, {Bailey}, {Barkhouser}, {Bautista}, {Beifiori}, {Berlind},
  {Bhardwaj}, {Bizyaev}, {Blake}, {Blanton}, {Blomqvist}, {Bolton}, {Borde},
  {Bovy}, {Brandt}, {Brewington}, {Brinkmann}, {Brown}, {Brownstein}, {Bundy},
  {Busca}, {Carithers}, {Carnero}, {Carr}, {Chen}, {Comparat}, {Connolly},
  {Cope}, {Croft}, {Cuesta}, {da Costa}, {Davenport}, {Delubac}, {de Putter},
  {Dhital}, {Ealet}, {Ebelke}, {Eisenstein}, {Escoffier}, {Fan}, {Filiz Ak},
  {Finley}, {Font-Ribera}, {G{\'e}nova-Santos}, {Gunn}, {Guo}, {Haggard},
  {Hall}, {Hamilton}, {Harris}, {Harris}, {Ho}, {Hogg}, {Holder}, {Honscheid},
  {Huehnerhoff}, {Jordan}, {Jordan}, {Kauffmann}, {Kazin}, {Kirkby}, {Klaene},
  {Kneib}, {Le Goff}, {Lee}, {Long}, {Loomis}, {Lundgren}, {Lupton}, {Maia},
  {Makler}, {Malanushenko}, {Malanushenko}, {Mandelbaum}, {Manera}, {Maraston},
  {Margala}, {Masters}, {McBride}, {McDonald}, {McGreer}, {McMahon}, {Mena},
  {Miralda-Escud{\'e}}, {Montero-Dorta}, {Montesano}, {Muna}, {Myers},
  {Naugle}, {Nichol}, {Noterdaeme}, {Nuza}, {Olmstead}, {Oravetz}, {Oravetz},
  {Owen}, {Padmanabhan}, {Palanque-Delabrouille}, {Pan}, {Parejko},
  {P{\^a}ris}, {Percival}, {P{\'e}rez-Fournon}, {P{\'e}rez-R{\`a}fols},
  {Petitjean}, {Pfaffenberger}, {Pforr}, {Pieri}, {Prada}, {Price-Whelan},
  {Raddick}, {Rebolo}, {Rich}, {Richards}, {Rockosi}, {Roe}, {Ross}, {Ross},
  {Rossi}, {Rubi{\~n}o-Martin}, {Samushia}, {S{\'a}nchez}, {Sayres}, {Schmidt},
  {Schneider}, {Sc{\'o}ccola}, {Seo}, {Shelden}, {Sheldon}, {Shen}, {Shu},
  {Slosar}, {Smee}, {Snedden}, {Stauffer}, {Steele}, {Strauss}, {Streblyanska},
  {Suzuki}, {Swanson}, {Tal}, {Tanaka}, {Thomas}, {Tinker}, {Tojeiro},
  {Tremonti}, {Vargas Maga{\~n}a}, {Verde}, {Viel}, {Wake}, {Watson}, {Weaver},
  {Weinberg}, {Weiner}, {West}, {White}, {Wood-Vasey}, {Yeche}, {Zehavi},
  {Zhao}, \& {Zheng}}]{Dawson2013}
{Dawson}, K.~S., {Schlegel}, D.~J., {Ahn}, C.~P., {et~al.} 2013, \aj, 145, 10,
  \dodoi{10.1088/0004-6256/145/1/10}

\bibitem[{{Dawson} {et~al.}(2016){Dawson}, {Kneib}, {Percival}, {Alam},
  {Albareti}, {Anderson}, {Armengaud}, {Aubourg}, {Bailey}, {Bautista},
  {Berlind}, {Bershady}, {Beutler}, {Bizyaev}, {Blanton}, {Blomqvist},
  {Bolton}, {Bovy}, {Brandt}, {Brinkmann}, {Brownstein}, {Burtin}, {Busca},
  {Cai}, {Chuang}, {Clerc}, {Comparat}, {Cope}, {Croft}, {Cruz-Gonzalez}, {da
  Costa}, {Cousinou}, {Darling}, {de la Macorra}, {de la Torre}, {Delubac}, {du
  Mas des Bourboux}, {Dwelly}, {Ealet}, {Eisenstein}, {Eracleous}, {Escoffier},
  {Fan}, {Finoguenov}, {Font-Ribera}, {Frinchaboy}, {Gaulme}, {Georgakakis},
  {Green}, {Guo}, {Guy}, {Ho}, {Holder}, {Huehnerhoff}, {Hutchinson}, {Jing},
  {Jullo}, {Kamble}, {Kinemuchi}, {Kirkby}, {Kitaura}, {Klaene}, {Laher},
  {Lang}, {Laurent}, {Le Goff}, {Li}, {Liang}, {Lima}, {Lin}, {Lin}, {Lin},
  {Long}, {Lundgren}, {MacDonald}, {Geimba Maia}, {Malanushenko},
  {Malanushenko}, {Mariappan}, {McBride}, {McGreer}, {M{\'e}nard}, {Merloni},
  {Meza}, {Montero-Dorta}, {Muna}, {Myers}, {Nandra}, {Naugle}, {Newman},
  {Noterdaeme}, {Nugent}, {Ogando}, {Olmstead}, {Oravetz}, {Oravetz},
  {Padmanabhan}, {Palanque-Delabrouille}, {Pan}, {Parejko}, {P{\^a}ris},
  {Peacock}, {Petitjean}, {Pieri}, {Pisani}, {Prada}, {Prakash}, {Raichoor},
  {Reid}, {Rich}, {Ridl}, {Rodriguez-Torres}, {Carnero Rosell}, {Ross},
  {Rossi}, {Ruan}, {Salvato}, {Sayres}, {Schneider}, {Schlegel}, {Seljak},
  {Seo}, {Sesar}, {Shandera}, {Shu}, {Slosar}, {Sobreira}, {Streblyanska},
  {Suzuki}, {Taylor}, {Tao}, {Tinker}, {Tojeiro}, {Vargas-Maga{\~n}a}, {Wang},
  {Weaver}, {Weinberg}, {White}, {Wood-Vasey}, {Yeche}, {Zhai}, {Zhao}, {Zhao},
  {Zheng}, {Ben Zhu}, \& {Zou}}]{Dawson2016}
{Dawson}, K.~S., {Kneib}, J.-P., {Percival}, W.~J., {et~al.} 2016, \aj, 151,
  44, \dodoi{10.3847/0004-6256/151/2/44}

\bibitem[{{DESI Collaboration} {et~al.}(2016{\natexlab{a}}){DESI
  Collaboration}, {Aghamousa}, {Aguilar}, {Ahlen}, {Alam}, {Allen}, {Allende
  Prieto}, {Annis}, {Bailey}, {Balland}, {Ballester}, {Baltay}, {Beaufore},
  {Bebek}, {Beers}, {Bell}, {Bernal}, {Besuner}, {Beutler}, {Blake}, {Bleuler},
  {Blomqvist}, {Blum}, {Bolton}, {Briceno}, {Brooks}, {Brownstein},
  {Buckley-Geer}, {Burden}, {Burtin}, {Busca}, {Cahn}, {Cai}, {Cardiel-Sas},
  {Carlberg}, {Carton}, {Casas}, {Castander}, {Cervantes-Cota}, {Claybaugh},
  {Close}, {Coker}, {Cole}, {Comparat}, {Cooper}, {Cousinou}, {Crocce}, {Cuby},
  {Cunningham}, {Davis}, {Dawson}, {de la Macorra}, {De Vicente}, {Delubac},
  {Derwent}, {Dey}, {Dhungana}, {Ding}, {Doel}, {Duan}, {Ealet}, {Edelstein},
  {Eftekharzadeh}, {Eisenstein}, {Elliott}, {Escoffier}, {Evatt}, {Fagrelius},
  {Fan}, {Fanning}, {Farahi}, {Farihi}, {Favole}, {Feng}, {Fernandez},
  {Findlay}, {Finkbeiner}, {Fitzpatrick}, {Flaugher}, {Flender}, {Font-Ribera},
  {Forero-Romero}, {Fosalba}, {Frenk}, {Fumagalli}, {Gaensicke}, {Gallo},
  {Garcia-Bellido}, {Gaztanaga}, {Pietro Gentile Fusillo}, {Gerard},
  {Gershkovich}, {Giannantonio}, {Gillet}, {Gonzalez-de-Rivera},
  {Gonzalez-Perez}, {Gott}, {Graur}, {Gutierrez}, {Guy}, {Habib}, {Heetderks},
  {Heetderks}, {Heitmann}, {Hellwing}, {Herrera}, {Ho}, {Holland}, {Honscheid},
  {Huff}, {Hutchinson}, {Huterer}, {Hwang}, {Illa Laguna}, {Ishikawa},
  {Jacobs}, {Jeffrey}, {Jelinsky}, {Jennings}, {Jiang}, {Jimenez}, {Johnson},
  {Joyce}, {Jullo}, {Juneau}, {Kama}, {Karcher}, {Karkar}, {Kehoe}, {Kennamer},
  {Kent}, {Kilbinger}, {Kim}, {Kirkby}, {Kisner}, {Kitanidis}, {Kneib},
  {Koposov}, {Kovacs}, {Koyama}, {Kremin}, {Kron}, {Kronig}, {Kueter-Young},
  {Lacey}, {Lafever}, {Lahav}, {Lambert}, {Lampton}, {Landriau}, {Lang},
  {Lauer}, {Le Goff}, {Le Guillou}, {Le Van Suu}, {Lee}, {Lee}, {Leitner},
  {Lesser}, {Levi}, {L'Huillier}, {Li}, {Liang}, {Lin}, {Linder}, {Loebman},
  {Luki{\'c}}, {Ma}, {MacCrann}, {Magneville}, {Makarem}, {Manera}, {Manser},
  {Marshall}, {Martini}, {Massey}, {Matheson}, {McCauley}, {McDonald},
  {McGreer}, {Meisner}, {Metcalfe}, {Miller}, {Miquel}, {Moustakas}, {Myers},
  {Naik}, {Newman}, {Nichol}, {Nicola}, {Nicolati da Costa}, {Nie}, {Niz},
  {Norberg}, {Nord}, {Norman}, {Nugent}, {O'Brien}, {Oh}, {Olsen}, {Padilla},
  {Padmanabhan}, {Padmanabhan}, {Palanque-Delabrouille}, {Palmese},
  {Pappalardo}, {P{\^a}ris}, {Park}, {Patej}, {Peacock}, {Peiris}, {Peng},
  {Percival}, {Perruchot}, {Pieri}, {Pogge}, {Pollack}, {Poppett}, {Prada},
  {Prakash}, {Probst}, {Rabinowitz}, {Raichoor}, {Ree}, {Refregier}, {Regal},
  {Reid}, {Reil}, {Rezaie}, {Rockosi}, {Roe}, {Ronayette}, {Roodman}, {Ross},
  {Ross}, {Rossi}, {Rozo}, {Ruhlmann-Kleider}, {Rykoff}, {Sabiu}, {Samushia},
  {Sanchez}, {Sanchez}, {Schlegel}, {Schneider}, {Schubnell}, {Secroun},
  {Seljak}, {Seo}, {Serrano}, {Shafieloo}, {Shan}, {Sharples}, {Sholl},
  {Shourt}, {Silber}, {Silva}, {Sirk}, {Slosar}, {Smith}, {Smoot}, {Som},
  {Song}, {Sprayberry}, {Staten}, {Stefanik}, {Tarle}, {Sien Tie}, {Tinker},
  {Tojeiro}, {Valdes}, {Valenzuela}, {Valluri}, {Vargas-Magana}, {Verde},
  {Walker}, {Wang}, {Wang}, {Weaver}, {Weaverdyck}, {Wechsler}, {Weinberg},
  {White}, {Yang}, {Yeche}, {Zhang}, {Zhao}, {Zheng}, {Zhou}, {Zhou}, {Zhu},
  {Zou}, \& {Zu}}]{DESI2016a}
{DESI Collaboration}, {Aghamousa}, A., {Aguilar}, J., {et~al.}
  2016{\natexlab{a}}, arXiv e-prints, arXiv:1611.00036.
\newblock \doarXiv{1611.00036}

\bibitem[{{DESI Collaboration} {et~al.}(2016{\natexlab{b}}){DESI
  Collaboration}, {Aghamousa}, {Aguilar}, {Ahlen}, {Alam}, {Allen}, {Allende
  Prieto}, {Annis}, {Bailey}, {Balland}, {Ballester}, {Baltay}, {Beaufore},
  {Bebek}, {Beers}, {Bell}, {Bernal}, {Besuner}, {Beutler}, {Blake}, {Bleuler},
  {Blomqvist}, {Blum}, {Bolton}, {Briceno}, {Brooks}, {Brownstein},
  {Buckley-Geer}, {Burden}, {Burtin}, {Busca}, {Cahn}, {Cai}, {Cardiel-Sas},
  {Carlberg}, {Carton}, {Casas}, {Castander}, {Cervantes-Cota}, {Claybaugh},
  {Close}, {Coker}, {Cole}, {Comparat}, {Cooper}, {Cousinou}, {Crocce}, {Cuby},
  {Cunningham}, {Davis}, {Dawson}, {de la Macorra}, {De Vicente}, {Delubac},
  {Derwent}, {Dey}, {Dhungana}, {Ding}, {Doel}, {Duan}, {Ealet}, {Edelstein},
  {Eftekharzadeh}, {Eisenstein}, {Elliott}, {Escoffier}, {Evatt}, {Fagrelius},
  {Fan}, {Fanning}, {Farahi}, {Farihi}, {Favole}, {Feng}, {Fernandez},
  {Findlay}, {Finkbeiner}, {Fitzpatrick}, {Flaugher}, {Flender}, {Font-Ribera},
  {Forero-Romero}, {Fosalba}, {Frenk}, {Fumagalli}, {Gaensicke}, {Gallo},
  {Garcia-Bellido}, {Gaztanaga}, {Pietro Gentile Fusillo}, {Gerard},
  {Gershkovich}, {Giannantonio}, {Gillet}, {Gonzalez-de-Rivera},
  {Gonzalez-Perez}, {Gott}, {Graur}, {Gutierrez}, {Guy}, {Habib}, {Heetderks},
  {Heetderks}, {Heitmann}, {Hellwing}, {Herrera}, {Ho}, {Holland}, {Honscheid},
  {Huff}, {Hutchinson}, {Huterer}, {Hwang}, {Illa Laguna}, {Ishikawa},
  {Jacobs}, {Jeffrey}, {Jelinsky}, {Jennings}, {Jiang}, {Jimenez}, {Johnson},
  {Joyce}, {Jullo}, {Juneau}, {Kama}, {Karcher}, {Karkar}, {Kehoe}, {Kennamer},
  {Kent}, {Kilbinger}, {Kim}, {Kirkby}, {Kisner}, {Kitanidis}, {Kneib},
  {Koposov}, {Kovacs}, {Koyama}, {Kremin}, {Kron}, {Kronig}, {Kueter-Young},
  {Lacey}, {Lafever}, {Lahav}, {Lambert}, {Lampton}, {Landriau}, {Lang},
  {Lauer}, {Le Goff}, {Le Guillou}, {Le Van Suu}, {Lee}, {Lee}, {Leitner},
  {Lesser}, {Levi}, {L'Huillier}, {Li}, {Liang}, {Lin}, {Linder}, {Loebman},
  {Luki{\'c}}, {Ma}, {MacCrann}, {Magneville}, {Makarem}, {Manera}, {Manser},
  {Marshall}, {Martini}, {Massey}, {Matheson}, {McCauley}, {McDonald},
  {McGreer}, {Meisner}, {Metcalfe}, {Miller}, {Miquel}, {Moustakas}, {Myers},
  {Naik}, {Newman}, {Nichol}, {Nicola}, {Nicolati da Costa}, {Nie}, {Niz},
  {Norberg}, {Nord}, {Norman}, {Nugent}, {O'Brien}, {Oh}, {Olsen}, {Padilla},
  {Padmanabhan}, {Padmanabhan}, {Palanque-Delabrouille}, {Palmese},
  {Pappalardo}, {P{\^a}ris}, {Park}, {Patej}, {Peacock}, {Peiris}, {Peng},
  {Percival}, {Perruchot}, {Pieri}, {Pogge}, {Pollack}, {Poppett}, {Prada},
  {Prakash}, {Probst}, {Rabinowitz}, {Raichoor}, {Ree}, {Refregier}, {Regal},
  {Reid}, {Reil}, {Rezaie}, {Rockosi}, {Roe}, {Ronayette}, {Roodman}, {Ross},
  {Ross}, {Rossi}, {Rozo}, {Ruhlmann-Kleider}, {Rykoff}, {Sabiu}, {Samushia},
  {Sanchez}, {Sanchez}, {Schlegel}, {Schneider}, {Schubnell}, {Secroun},
  {Seljak}, {Seo}, {Serrano}, {Shafieloo}, {Shan}, {Sharples}, {Sholl},
  {Shourt}, {Silber}, {Silva}, {Sirk}, {Slosar}, {Smith}, {Smoot}, {Som},
  {Song}, {Sprayberry}, {Staten}, {Stefanik}, {Tarle}, {Sien Tie}, {Tinker},
  {Tojeiro}, {Valdes}, {Valenzuela}, {Valluri}, {Vargas-Magana}, {Verde},
  {Walker}, {Wang}, {Wang}, {Weaver}, {Weaverdyck}, {Wechsler}, {Weinberg},
  {White}, {Yang}, {Yeche}, {Zhang}, {Zhao}, {Zheng}, {Zhou}, {Zhou}, {Zhu},
  {Zou}, \& {Zu}}]{DESI2016b}
---. 2016{\natexlab{b}}, arXiv e-prints, arXiv:1611.00037.
\newblock \doarXiv{1611.00037}

\bibitem[{{Dey} {et~al.}(2019){Dey}, {Schlegel}, {Lang}, {Blum}, {Burleigh},
  {Fan}, {Findlay}, {Finkbeiner}, {Herrera}, {Juneau}, {Landriau}, {Levi},
  {McGreer}, {Meisner}, {Myers}, {Moustakas}, {Nugent}, {Patej}, {Schlafly},
  {Walker}, {Valdes}, {Weaver}, {Y{\`e}che}, {Zou}, {Zhou}, {Abareshi},
  {Abbott}, {Abolfathi}, {Aguilera}, {Alam}, {Allen}, {Alvarez}, {Annis},
  {Ansarinejad}, {Aubert}, {Beechert}, {Bell}, {BenZvi}, {Beutler}, {Bielby},
  {Bolton}, {Brice{\~n}o}, {Buckley-Geer}, {Butler}, {Calamida}, {Carlberg},
  {Carter}, {Casas}, {Castander}, {Choi}, {Comparat}, {Cukanovaite}, {Delubac},
  {DeVries}, {Dey}, {Dhungana}, {Dickinson}, {Ding}, {Donaldson}, {Duan},
  {Duckworth}, {Eftekharzadeh}, {Eisenstein}, {Etourneau}, {Fagrelius},
  {Farihi}, {Fitzpatrick}, {Font-Ribera}, {Fulmer}, {G{\"a}nsicke},
  {Gaztanaga}, {George}, {Gerdes}, {Gontcho}, {Gorgoni}, {Green}, {Guy},
  {Harmer}, {Hernandez}, {Honscheid}, {Huang}, {James}, {Jannuzi}, {Jiang},
  {Joyce}, {Karcher}, {Karkar}, {Kehoe}, {Kneib}, {Kueter-Young}, {Lan},
  {Lauer}, {Le Guillou}, {Le Van Suu}, {Lee}, {Lesser}, {Perreault Levasseur},
  {Li}, {Mann}, {Marshall}, {Mart{\'\i}nez-V{\'a}zquez}, {Martini}, {du Mas des
  Bourboux}, {McManus}, {Meier}, {M{\'e}nard}, {Metcalfe},
  {Mu{\~n}oz-Guti{\'e}rrez}, {Najita}, {Napier}, {Narayan}, {Newman}, {Nie},
  {Nord}, {Norman}, {Olsen}, {Paat}, {Palanque-Delabrouille}, {Peng},
  {Poppett}, {Poremba}, {Prakash}, {Rabinowitz}, {Raichoor}, {Rezaie},
  {Robertson}, {Roe}, {Ross}, {Ross}, {Rudnick}, {Safonova}, {Saha},
  {S{\'a}nchez}, {Savary}, {Schweiker}, {Scott}, {Seo}, {Shan}, {Silva},
  {Slepian}, {Soto}, {Sprayberry}, {Staten}, {Stillman}, {Stupak}, {Summers},
  {Sien Tie}, {Tirado}, {Vargas-Maga{\~n}a}, {Vivas}, {Wechsler}, {Williams},
  {Yang}, {Yang}, {Yapici}, {Zaritsky}, {Zenteno}, {Zhang}, {Zhang}, {Zhou}, \&
  {Zhou}}]{Dey2019}
{Dey}, A., {Schlegel}, D.~J., {Lang}, D., {et~al.} 2019, \aj, 157, 168,
  \dodoi{10.3847/1538-3881/ab089d}

\bibitem[{{Diamond-Stanic} {et~al.}(2021){Diamond-Stanic}, {Moustakas}, {Sell},
  {Tremonti}, {Coil}, {Davis}, {Geach}, {Gottlieb}, {Hickox}, {Kepley},
  {Lipscomb}, {Rines}, {Rudnick}, {Thompson}, {Valdez}, {Bradna}, {Camarillo},
  {Cinquino}, {Ohene}, {Perrotta}, {Petter}, {Rupke}, {Umeh}, \&
  {Whalen}}]{Diamond-Stanic2021}
{Diamond-Stanic}, A.~M., {Moustakas}, J., {Sell}, P.~H., {et~al.} 2021, \apj,
  912, 11, \dodoi{10.3847/1538-4357/abe935}

\bibitem[{{D{\'\i}az-Garc{\'\i}a} {et~al.}(2019){D{\'\i}az-Garc{\'\i}a},
  {Cenarro}, {L{\'o}pez-Sanjuan}, {Ferreras}, {Fern{\'a}ndez-Soto},
  {Gonz{\'a}lez Delgado}, {M{\'a}rquez}, {Masegosa}, {San Roman}, {Viironen},
  {Bonoli}, {Cervi{\~n}o}, {Moles}, {Crist{\'o}bal-Hornillos}, {Alfaro},
  {Aparicio-Villegas}, {Ben{\'\i}tez}, {Broadhurst}, {Cabrera-Ca{\~n}o},
  {Castander}, {Cepa}, {Husillos}, {Infante}, {Aguerri}, {Mart{\'\i}nez},
  {Molino}, {del Olmo}, {Perea}, {Prada}, \& {Quintana}}]{Diaz-Garcia2019}
{D{\'\i}az-Garc{\'\i}a}, L.~A., {Cenarro}, A.~J., {L{\'o}pez-Sanjuan}, C.,
  {et~al.} 2019, \aap, 631, A157, \dodoi{10.1051/0004-6361/201832882}

\bibitem[{{Donnari} {et~al.}(2019){Donnari}, {Pillepich}, {Nelson},
  {Vogelsberger}, {Genel}, {Weinberger}, {Marinacci}, {Springel}, \&
  {Hernquist}}]{Donnari2019}
{Donnari}, M., {Pillepich}, A., {Nelson}, D., {et~al.} 2019, \mnras, 485, 4817,
  \dodoi{10.1093/mnras/stz712}

\bibitem[{{Dotter}(2016)}]{Dotter2016}
{Dotter}, A. 2016, \apjs, 222, 8, \dodoi{10.3847/0067-0049/222/1/8}

\bibitem[{{Draine} {et~al.}(2007){Draine}, {Dale}, {Bendo}, {Gordon}, {Smith},
  {Armus}, {Engelbracht}, {Helou}, {Kennicutt}, {Li}, {Roussel}, {Walter},
  {Calzetti}, {Moustakas}, {Murphy}, {Rieke}, {Bot}, {Hollenbach}, {Sheth}, \&
  {Teplitz}}]{Draine2007}
{Draine}, B.~T., {Dale}, D.~A., {Bendo}, G., {et~al.} 2007, \apj, 663, 866,
  \dodoi{10.1086/518306}

\bibitem[{{Dressler} \& {Gunn}(1983)}]{Dressler1983}
{Dressler}, A., \& {Gunn}, J.~E. 1983, \apj, 270, 7, \dodoi{10.1086/161093}

\bibitem[{{Dressler} {et~al.}(2018){Dressler}, {Kelson}, \&
  {Abramson}}]{Dressler2018}
{Dressler}, A., {Kelson}, D.~D., \& {Abramson}, L.~E. 2018, \apj, 869, 152,
  \dodoi{10.3847/1538-4357/aaedbe}

\bibitem[{{Dressler} {et~al.}(2004){Dressler}, {Oemler}, {Poggianti}, {Smail},
  {Trager}, {Shectman}, {Couch}, \& {Ellis}}]{Dressler2004}
{Dressler}, A., {Oemler}, Augustus, J., {Poggianti}, B.~M., {et~al.} 2004,
  \apj, 617, 867, \dodoi{10.1086/424890}

\bibitem[{{Dressler} {et~al.}(2016){Dressler}, {Kelson}, {Abramson},
  {Gladders}, {Oemler}, {Poggianti}, {Mulchaey}, {Vulcani}, {Shectman},
  {Williams}, \& {McCarthy}}]{Dressler2016}
{Dressler}, A., {Kelson}, D.~D., {Abramson}, L.~E., {et~al.} 2016, \apj, 833,
  251, \dodoi{10.3847/1538-4357/833/2/251}

\bibitem[{{Eisenstein} {et~al.}(2001){Eisenstein}, {Annis}, {Gunn}, {Szalay},
  {Connolly}, {Nichol}, {Bahcall}, {Bernardi}, {Burles}, {Castander},
  {Fukugita}, {Hogg}, {Ivezi{\'c}}, {Knapp}, {Lupton}, {Narayanan}, {Postman},
  {Reichart}, {Richmond}, {Schneider}, {Schlegel}, {Strauss}, {SubbaRao},
  {Tucker}, {Vanden Berk}, {Vogeley}, {Weinberg}, \& {Yanny}}]{Eisenstein2001}
{Eisenstein}, D.~J., {Annis}, J., {Gunn}, J.~E., {et~al.} 2001, \aj, 122, 2267,
  \dodoi{10.1086/323717}

\bibitem[{{Ellison} {et~al.}(2022){Ellison}, {Wilkinson}, {Woo}, {Leung},
  {Wild}, {Bickley}, {Patton}, {Quai}, \& {Gwyn}}]{Ellison2022}
{Ellison}, S.~L., {Wilkinson}, S., {Woo}, J., {et~al.} 2022, arXiv e-prints,
  arXiv:2209.07613.
\newblock \doarXiv{2209.07613}

\bibitem[{{Estrada-Carpenter} {et~al.}(2019){Estrada-Carpenter}, {Papovich},
  {Momcheva}, {Brammer}, {Long}, {Quadri}, {Bridge}, {Dickinson}, {Ferguson},
  {Finkelstein}, {Giavalisco}, {Gosmeyer}, {Lotz}, {Salmon}, {Skelton},
  {Trump}, \& {Weiner}}]{Estrada-Carpenter2019}
{Estrada-Carpenter}, V., {Papovich}, C., {Momcheva}, I., {et~al.} 2019, \apj,
  870, 133, \dodoi{10.3847/1538-4357/aaf22e}

\bibitem[{{Falc{\'o}n-Barroso} {et~al.}(2011){Falc{\'o}n-Barroso},
  {S{\'a}nchez-Bl{\'a}zquez}, {Vazdekis}, {Ricciardelli}, {Cardiel}, {Cenarro},
  {Gorgas}, \& {Peletier}}]{Falcon-Barroso2011}
{Falc{\'o}n-Barroso}, J., {S{\'a}nchez-Bl{\'a}zquez}, P., {Vazdekis}, A.,
  {et~al.} 2011, \aap, 532, A95, \dodoi{10.1051/0004-6361/201116842}

\bibitem[{{French}(2021)}]{French2021}
{French}, K.~D. 2021, \pasp, 133, 072001, \dodoi{10.1088/1538-3873/ac0a59}

\bibitem[{{French} {et~al.}(2015){French}, {Yang}, {Zabludoff}, {Narayanan},
  {Shirley}, {Walter}, {Smith}, \& {Tremonti}}]{French2015}
{French}, K.~D., {Yang}, Y., {Zabludoff}, A., {et~al.} 2015, \apj, 801, 1,
  \dodoi{10.1088/0004-637X/801/1/1}

\bibitem[{{French} {et~al.}(2018){French}, {Yang}, {Zabludoff}, \&
  {Tremonti}}]{French2018a}
{French}, K.~D., {Yang}, Y., {Zabludoff}, A.~I., \& {Tremonti}, C.~A. 2018,
  \apj, 862, 2, \dodoi{10.3847/1538-4357/aacb2d}

\bibitem[{{Fumagalli} {et~al.}(2016){Fumagalli}, {Franx}, {van Dokkum},
  {Whitaker}, {Skelton}, {Brammer}, {Nelson}, {Maseda}, {Momcheva}, {Kriek},
  {Labb{\'e}}, {Lundgren}, \& {Rix}}]{Fumagalli2016}
{Fumagalli}, M., {Franx}, M., {van Dokkum}, P., {et~al.} 2016, \apj, 822, 1,
  \dodoi{10.3847/0004-637X/822/1/1}

\bibitem[{{Gallazzi} {et~al.}(2014){Gallazzi}, {Bell}, {Zibetti}, {Brinchmann},
  \& {Kelson}}]{Gallazzi2014}
{Gallazzi}, A., {Bell}, E.~F., {Zibetti}, S., {Brinchmann}, J., \& {Kelson},
  D.~D. 2014, \apj, 788, 72, \dodoi{10.1088/0004-637X/788/1/72}

\bibitem[{{Gallazzi} {et~al.}(2005){Gallazzi}, {Charlot}, {Brinchmann},
  {White}, \& {Tremonti}}]{Gallazzi2005}
{Gallazzi}, A., {Charlot}, S., {Brinchmann}, J., {White}, S. D.~M., \&
  {Tremonti}, C.~A. 2005, \mnras, 362, 41,
  \dodoi{10.1111/j.1365-2966.2005.09321.x}

\bibitem[{{Greene} {et~al.}(2022){Greene}, {Bezanson}, {Ouchi}, {Silverman}, \&
  {the PFS Galaxy Evolution Working Group}}]{Greene2022}
{Greene}, J., {Bezanson}, R., {Ouchi}, M., {Silverman}, J., \& {the PFS Galaxy
  Evolution Working Group}. 2022, arXiv e-prints, arXiv:2206.14908.
\newblock \doarXiv{2206.14908}

\bibitem[{{Greene} {et~al.}(2020){Greene}, {Setton}, {Bezanson}, {Suess},
  {Kriek}, {Spilker}, {Goulding}, \& {Feldmann}}]{Greene2020}
{Greene}, J.~E., {Setton}, D., {Bezanson}, R., {et~al.} 2020, \apjl, 899, L9,
  \dodoi{10.3847/2041-8213/aba534}

\bibitem[{{Guy} {et~al.}(2022){Guy}, {Bailey}, {Kremin}, {Alam}, {Allende
  Prieto}, {BenZvi}, {Bolton}, {Brooks}, {Chaussidon}, {Cooper}, {Dawson}, {de
  la Macorra}, {Dey}, {Dey}, {Dhungana}, {Eisenstein}, {Font-Ribera},
  {Forero-Romero}, {Gazta{\~n}aga}, {Gontcho}, {Green}, {Honscheid}, {Ishak},
  {Kehoe}, {Kirkby}, {Kisner}, {Koposov}, {Lan}, {Landriau}, {Le Guillou},
  {Levi}, {Magneville}, {Manser}, {Martini}, {Meisner}, {Miquel}, {Moustakas},
  {Myers}, {Newman}, {Nie}, {Palanque-Delabrouille}, {Percival}, {Poppett},
  {Prada}, {Raichoor}, {Ravoux}, {Ross}, {Schlafly}, {Schlegel}, {Schubnell},
  {Sharples}, {Tarl{\'e}}, {Weaver}, {Y{\`e}che}, {Zhou}, {Zhou}, \&
  {Zou}}]{Guy2022}
{Guy}, J., {Bailey}, S., {Kremin}, A., {et~al.} 2022, arXiv e-prints,
  arXiv:2209.14482.
\newblock \doarXiv{2209.14482}

\bibitem[{{Hopkins} {et~al.}(2008){Hopkins}, {Cox}, {Kere{\v{s}}}, \&
  {Hernquist}}]{Hopkins2008}
{Hopkins}, P.~F., {Cox}, T.~J., {Kere{\v{s}}}, D., \& {Hernquist}, L. 2008,
  \apjs, 175, 390, \dodoi{10.1086/524363}

\bibitem[{{Johnson} \& {Leja}(2017)}]{Johnson2017}
{Johnson}, B., \& {Leja}, J. 2017, {Bd-J/Prospector: Initial Release}, v0.1,
  Zenodo, \dodoi{10.5281/zenodo.1116491}

\bibitem[{{Johnson} {et~al.}(2021){Johnson}, {Foreman-Mackey}, {Sick}, {Leja},
  {Byler}, {Walmsley}, {Tollerud}, {Leung}, \& {Scott}}]{Johnson2021}
{Johnson}, B., {Foreman-Mackey}, D., {Sick}, J., {et~al.} 2021,
  {dfm/python-fsps: python-fsps v0.4.0}, v0.4.0,  Zenodo,
  \dodoi{10.5281/zenodo.4577191}

\bibitem[{{Juneau} {et~al.}(2005){Juneau}, {Glazebrook}, {Crampton},
  {McCarthy}, {Savaglio}, {Abraham}, {Carlberg}, {Chen}, {Le Borgne}, {Marzke},
  {Roth}, {J{\o}rgensen}, {Hook}, \& {Murowinski}}]{Juneau2005}
{Juneau}, S., {Glazebrook}, K., {Crampton}, D., {et~al.} 2005, \apjl, 619,
  L135, \dodoi{10.1086/427937}

\bibitem[{{Khullar} {et~al.}(2022){Khullar}, {Bayliss}, {Gladders}, {Kim},
  {Calzadilla}, {Strazzullo}, {Bleem}, {Mahler}, {McDonald}, {Floyd},
  {Reichardt}, {Ruppin}, {Saro}, {Sharon}, {Somboonpanyakul}, {Stalder}, \&
  {Stark}}]{Khullar2022}
{Khullar}, G., {Bayliss}, M.~B., {Gladders}, M.~D., {et~al.} 2022, \apj, 934,
  177, \dodoi{10.3847/1538-4357/ac7c0c}

\bibitem[{{Kriek} \& {Conroy}(2013)}]{Kriek2013}
{Kriek}, M., \& {Conroy}, C. 2013, \apjl, 775, L16,
  \dodoi{10.1088/2041-8205/775/1/L16}

\bibitem[{{Leja} {et~al.}(2017){Leja}, {Johnson}, {Conroy}, {van Dokkum}, \&
  {Byler}}]{Leja2017}
{Leja}, J., {Johnson}, B.~D., {Conroy}, C., {van Dokkum}, P.~G., \& {Byler}, N.
  2017, \apj, 837, 170, \dodoi{10.3847/1538-4357/aa5ffe}

\bibitem[{{Leja} {et~al.}(2020){Leja}, {Speagle}, {Johnson}, {Conroy}, {van
  Dokkum}, \& {Franx}}]{Leja2020}
{Leja}, J., {Speagle}, J.~S., {Johnson}, B.~D., {et~al.} 2020, \apj, 893, 111,
  \dodoi{10.3847/1538-4357/ab7e27}

\bibitem[{{Leja} {et~al.}(2019){Leja}, {Johnson}, {Conroy}, {van Dokkum},
  {Speagle}, {Brammer}, {Momcheva}, {Skelton}, {Whitaker}, {Franx}, \&
  {Nelson}}]{Leja2019}
{Leja}, J., {Johnson}, B.~D., {Conroy}, C., {et~al.} 2019, \apj, 877, 140,
  \dodoi{10.3847/1538-4357/ab1d5a}

\bibitem[{{Leja} {et~al.}(2021){Leja}, {Speagle}, {Ting}, {Johnson}, {Conroy},
  {Whitaker}, {Nelson}, {van Dokkum}, \& {Franx}}]{Leja2021}
{Leja}, J., {Speagle}, J.~S., {Ting}, Y.-S., {et~al.} 2021, arXiv e-prints,
  arXiv:2110.04314.
\newblock \doarXiv{2110.04314}

\bibitem[{{Levi} {et~al.}(2013){Levi}, {Bebek}, {Beers}, {Blum}, {Cahn},
  {Eisenstein}, {Flaugher}, {Honscheid}, {Kron}, {Lahav}, {McDonald}, {Roe},
  {Schlegel}, \& {representing the DESI collaboration}}]{Levi2013}
{Levi}, M., {Bebek}, C., {Beers}, T., {et~al.} 2013, arXiv e-prints,
  arXiv:1308.0847.
\newblock \doarXiv{1308.0847}

\bibitem[{{Maiolino} {et~al.}(2020){Maiolino}, {Cirasuolo}, {Afonso}, {Bauer},
  {Bowler}, {Cucciati}, {Daddi}, {De Lucia}, {Evans}, {Flores}, {Gargiulo},
  {Garilli}, {Jablonka}, {Jarvis}, {Kneib}, {Lilly}, {Looser}, {Magliocchetti},
  {Man}, {Mannucci}, {Maurogordato}, {McLure}, {Norberg}, {Oesch}, {Oliva},
  {Paltani}, {Pappalardo}, {Peng}, {Pentericci}, {Pozzetti}, {Renzini},
  {Rodrigues}, {Royer}, {Serjeant}, {Vanzi}, {Wild}, \&
  {Zamorani}}]{Maiolino2020}
{Maiolino}, R., {Cirasuolo}, M., {Afonso}, J., {et~al.} 2020, The Messenger,
  180, 24, \dodoi{10.18727/0722-6691/5197}

\bibitem[{{Maltby} {et~al.}(2018){Maltby}, {Almaini}, {Wild}, {Hatch},
  {Hartley}, {Simpson}, {Rowlands}, \& {Socolovsky}}]{Maltby2018}
{Maltby}, D.~T., {Almaini}, O., {Wild}, V., {et~al.} 2018, \mnras, 480, 381,
  \dodoi{10.1093/mnras/sty1794}

\bibitem[{{McLure} {et~al.}(2018){McLure}, {Pentericci}, {Cimatti}, {Dunlop},
  {Elbaz}, {Fontana}, {Nandra}, {Amorin}, {Bolzonella}, {Bongiorno}, {Carnall},
  {Castellano}, {Cirasuolo}, {Cucciati}, {Cullen}, {De Barros}, {Finkelstein},
  {Fontanot}, {Franzetti}, {Fumana}, {Gargiulo}, {Garilli}, {Guaita},
  {Hartley}, {Iovino}, {Jarvis}, {Juneau}, {Karman}, {Maccagni}, {Marchi},
  {M{\'a}rmol-Queralt{\'o}}, {Pompei}, {Pozzetti}, {Scodeggio}, {Sommariva},
  {Talia}, {Almaini}, {Balestra}, {Bardelli}, {Bell}, {Bourne}, {Bowler},
  {Brusa}, {Buitrago}, {Caputi}, {Cassata}, {Charlot}, {Citro}, {Cresci},
  {Cristiani}, {Curtis-Lake}, {Dickinson}, {Fazio}, {Ferguson}, {Fiore},
  {Franco}, {Fynbo}, {Galametz}, {Georgakakis}, {Giavalisco}, {Grazian},
  {Hathi}, {Jung}, {Kim}, {Koekemoer}, {Khusanova}, {Le F{\`e}vre}, {Lotz},
  {Mannucci}, {Maltby}, {Matsuoka}, {McLeod}, {Mendez-Hernandez},
  {Mendez-Abreu}, {Mignoli}, {Moresco}, {Mortlock}, {Nonino}, {Pannella},
  {Papovich}, {Popesso}, {Rosario}, {Salvato}, {Santini}, {Schaerer},
  {Schreiber}, {Stark}, {Tasca}, {Thomas}, {Treu}, {Vanzella}, {Wild},
  {Williams}, {Zamorani}, \& {Zucca}}]{McLure2018}
{McLure}, R.~J., {Pentericci}, L., {Cimatti}, A., {et~al.} 2018, \mnras, 479,
  25, \dodoi{10.1093/mnras/sty1213}

\bibitem[{{Muzzin} {et~al.}(2013){Muzzin}, {Marchesini}, {Stefanon}, {Franx},
  {McCracken}, {Milvang-Jensen}, {Dunlop}, {Fynbo}, {Brammer}, {Labb{\'e}}, \&
  {van Dokkum}}]{Muzzin2013}
{Muzzin}, A., {Marchesini}, D., {Stefanon}, M., {et~al.} 2013, \apj, 777, 18,
  \dodoi{10.1088/0004-637X/777/1/18}

\bibitem[{{Pacifici} {et~al.}(2016){Pacifici}, {Kassin}, {Weiner}, {Holden},
  {Gardner}, {Faber}, {Ferguson}, {Koo}, {Primack}, {Bell}, {Dekel}, {Gawiser},
  {Giavalisco}, {Rafelski}, {Simons}, {Barro}, {Croton}, {Dav{\'e}}, {Fontana},
  {Grogin}, {Koekemoer}, {Lee}, {Salmon}, {Somerville}, \&
  {Behroozi}}]{Pacifici2016}
{Pacifici}, C., {Kassin}, S.~A., {Weiner}, B.~J., {et~al.} 2016, \apj, 832, 79,
  \dodoi{10.3847/0004-637X/832/1/79}

\bibitem[{{Park} {et~al.}(2022){Park}, {Belli}, {Conroy}, {Tacchella}, {Leja},
  {Cutler}, {Johnson}, {Nelson}, \& {Emami}}]{Park2022}
{Park}, M., {Belli}, S., {Conroy}, C., {et~al.} 2022, arXiv e-prints,
  arXiv:2210.03747.
\newblock \doarXiv{2210.03747}

\bibitem[{{Patel} {et~al.}(2011){Patel}, {Kelson}, {Holden}, {Franx}, \&
  {Illingworth}}]{Patel2011}
{Patel}, S.~G., {Kelson}, D.~D., {Holden}, B.~P., {Franx}, M., \&
  {Illingworth}, G.~D. 2011, \apj, 735, 53, \dodoi{10.1088/0004-637X/735/1/53}

\bibitem[{{Pattarakijwanich} {et~al.}(2016){Pattarakijwanich}, {Strauss}, {Ho},
  \& {Ross}}]{Pattarakijwanich2016}
{Pattarakijwanich}, P., {Strauss}, M.~A., {Ho}, S., \& {Ross}, N.~P. 2016,
  \apj, 833, 19, \dodoi{10.3847/0004-637X/833/1/19}

\bibitem[{{Pawlik} {et~al.}(2016){Pawlik}, {Wild}, {Walcher}, {Johansson},
  {Villforth}, {Rowlands}, {Mendez-Abreu}, \& {Hewlett}}]{Pawlik2016}
{Pawlik}, M.~M., {Wild}, V., {Walcher}, C.~J., {et~al.} 2016, \mnras, 456,
  3032, \dodoi{10.1093/mnras/stv2878}

\bibitem[{{Quai} {et~al.}(2021){Quai}, {Hani}, {Ellison}, {Patton}, \&
  {Woo}}]{Quai2021}
{Quai}, S., {Hani}, M.~H., {Ellison}, S.~L., {Patton}, D.~R., \& {Woo}, J.
  2021, \mnras, 504, 1888, \dodoi{10.1093/mnras/stab988}

\bibitem[{{Rowlands} {et~al.}(2018){Rowlands}, {Heckman}, {Wild}, {Zakamska},
  {Rodriguez-Gomez}, {Barrera-Ballesteros}, {Lotz}, {Thilker}, {Andrews},
  {Boquien}, {Brinkmann}, {Brownstein}, {Hwang}, \&
  {Smethurst}}]{Rowlands2018b}
{Rowlands}, K., {Heckman}, T., {Wild}, V., {et~al.} 2018, \mnras, 480, 2544,
  \dodoi{10.1093/mnras/sty1916}

\bibitem[{{S{\'a}nchez-Bl{\'a}zquez} {et~al.}(2006){S{\'a}nchez-Bl{\'a}zquez},
  {Peletier}, {Jim{\'e}nez-Vicente}, {Cardiel}, {Cenarro},
  {Falc{\'o}n-Barroso}, {Gorgas}, {Selam}, \&
  {Vazdekis}}]{Sanchez-Blazquez2006}
{S{\'a}nchez-Bl{\'a}zquez}, P., {Peletier}, R.~F., {Jim{\'e}nez-Vicente}, J.,
  {et~al.} 2006, \mnras, 371, 703, \dodoi{10.1111/j.1365-2966.2006.10699.x}

\bibitem[{{Sazonova} {et~al.}(2021){Sazonova}, {Alatalo}, {Rowlands},
  {Deustua}, {French}, {Heckman}, {Lanz}, {Lisenfeld}, {Luo}, {Medling},
  {Nyland}, {Otter}, {Petric}, {Snyder}, \& {Urry}}]{Sazonova2021}
{Sazonova}, E., {Alatalo}, K., {Rowlands}, K., {et~al.} 2021, \apj, 919, 134,
  \dodoi{10.3847/1538-4357/ac0f7f}

\bibitem[{{Schawinski} {et~al.}(2014){Schawinski}, {Urry}, {Simmons},
  {Fortson}, {Kaviraj}, {Keel}, {Lintott}, {Masters}, {Nichol}, {Sarzi},
  {Skibba}, {Treister}, {Willett}, {Wong}, \& {Yi}}]{Schawinski2014}
{Schawinski}, K., {Urry}, C.~M., {Simmons}, B.~D., {et~al.} 2014, \mnras, 440,
  889, \dodoi{10.1093/mnras/stu327}

\bibitem[{{Schlegel} {et~al.}(1998){Schlegel}, {Finkbeiner}, \&
  {Davis}}]{Schlegel1998}
{Schlegel}, D.~J., {Finkbeiner}, D.~P., \& {Davis}, M. 1998, \apj, 500, 525,
  \dodoi{10.1086/305772}

\bibitem[{{Setton} {et~al.}(2022){Setton}, {Verrico}, {Bezanson}, {Greene},
  {Suess}, {Goulding}, {Spilker}, {Kriek}, {Feldmann}, {Narayanan},
  {Hall-Hooper}, \& {Kado-Fong}}]{Setton2022a}
{Setton}, D.~J., {Verrico}, M., {Bezanson}, R., {et~al.} 2022, \apj, 931, 51,
  \dodoi{10.3847/1538-4357/ac6096}

\bibitem[{{Silber} {et~al.}(2022){Silber}, {Fagrelius}, {Fanning}, {Schubnell},
  {Aguilar}, {Ahlen}, {Ameel}, {Ballester}, {Baltay}, {Bebek}, {Beard},
  {Besuner}, {Cardiel-Sas}, {Casas}, {Castander}, {Claybaugh}, {Dobson},
  {Duan}, {Dunlop}, {Edelstein}, {Emmet}, {Elliott}, {Evatt}, {Gershkovich},
  {Guy}, {Harris}, {Heetderks}, {Heetderks}, {Honscheid}, {Illa}, {Jelinsky},
  {Jelinsky}, {Jimenez}, {Karcher}, {Kent}, {Kirkby}, {Kneib}, {Lambert},
  {Lampton}, {Leitner}, {Levi}, {McCauley}, {Meisner}, {Miller}, {Miquel},
  {Mundet}, {Poppett}, {Rabinowitz}, {Reil}, {Roman}, {Schlegel}, {Serrano},
  {Van Shourt}, {Sprayberry}, {Tarl{\'e}}, {Sien Tie}, {Weaverdyck}, {Zhang},
  {Azzaro}, {Bailey}, {Becerril}, {Blackwell}, {Bouri}, {Brooks},
  {Buckley-Geer}, {Pe{\~n}ate Castro}, {Derwent}, {Dey}, {Dhungana}, {Doel},
  {Eisenstein}, {Fahim}, {Garcia-Bellido}, {Gazta{\~n}aga}, {Gontcho},
  {Gutierrez}, {H{\"o}rler}, {Kehoe}, {Kisner}, {Kremin}, {Kronig}, {Landriau},
  {Le Guillou}, {Martini}, {Moustakas}, {Palanque-Delabrouille}, {Peng},
  {Percival}, {Prada}, {Allende Prieto}, {Gonzalez de Rivera}, {Sanchez},
  {Sanchez}, {Sharples}, {Soares-Santos}, {Schlafly}, {Weaver}, {Zhou}, {Zhu},
  \& {Zou}}]{Silber2022}
{Silber}, J.~H., {Fagrelius}, P., {Fanning}, K., {et~al.} 2022, arXiv e-prints,
  arXiv:2205.09014.
\newblock \doarXiv{2205.09014}

\bibitem[{{Speagle}(2020)}]{Speagle2020}
{Speagle}, J.~S. 2020, \mnras, 493, 3132, \dodoi{10.1093/mnras/staa278}

\bibitem[{{Spilker} {et~al.}(2022){Spilker}, {Suess}, {Setton}, {Bezanson},
  {Feldmann}, {Greene}, {Kriek}, {Lower}, {Narayanan}, \&
  {Verrico}}]{Spilker2022}
{Spilker}, J.~S., {Suess}, K.~A., {Setton}, D.~J., {et~al.} 2022, \apjl, 936,
  L11, \dodoi{10.3847/2041-8213/ac75ea}

\bibitem[{{Suess} {et~al.}(2021){Suess}, {Kriek}, {Price}, \&
  {Barro}}]{Suess2021}
{Suess}, K.~A., {Kriek}, M., {Price}, S.~H., \& {Barro}, G. 2021, \apj, 915,
  87, \dodoi{10.3847/1538-4357/abf1e4}

\bibitem[{{Suess} {et~al.}(2022{\natexlab{a}}){Suess}, {Leja}, {Johnson},
  {Bezanson}, {Greene}, {Kriek}, {Lower}, {Narayanan}, {Setton}, \&
  {Spilker}}]{Suess2022b}
{Suess}, K.~A., {Leja}, J., {Johnson}, B.~D., {et~al.} 2022{\natexlab{a}},
  \apj, 935, 146, \dodoi{10.3847/1538-4357/ac82b0}

\bibitem[{{Suess} {et~al.}(2022{\natexlab{b}}){Suess}, {Kriek}, {Bezanson},
  {Greene}, {Setton}, {Spilker}, {Feldmann}, {Goulding}, {Johnson}, {Leja},
  {Narayanan}, {Hall-Hooper}, {Hunt}, {Lower}, \& {Verrico}}]{Suess2022a}
{Suess}, K.~A., {Kriek}, M., {Bezanson}, R., {et~al.} 2022{\natexlab{b}}, \apj,
  926, 89, \dodoi{10.3847/1538-4357/ac404a}

\bibitem[{{Tacchella} {et~al.}(2022){Tacchella}, {Conroy}, {Faber}, {Johnson},
  {Leja}, {Barro}, {Cunningham}, {Deason}, {Guhathakurta}, {Guo}, {Hernquist},
  {Koo}, {McKinnon}, {Rockosi}, {Speagle}, {van Dokkum}, \&
  {Yesuf}}]{Tacchella2022}
{Tacchella}, S., {Conroy}, C., {Faber}, S.~M., {et~al.} 2022, \apj, 926, 134,
  \dodoi{10.3847/1538-4357/ac449b}

\bibitem[{{Tinsley} \& {Gunn}(1976)}]{Tinsley1976}
{Tinsley}, B.~M., \& {Gunn}, J.~E. 1976, \apj, 203, 52, \dodoi{10.1086/154046}

\bibitem[{{Tremonti} {et~al.}(2007){Tremonti}, {Moustakas}, \&
  {Diamond-Stanic}}]{Tremonti2007}
{Tremonti}, C.~A., {Moustakas}, J., \& {Diamond-Stanic}, A.~M. 2007, \apjl,
  663, L77, \dodoi{10.1086/520083}

\bibitem[{{van der Wel} {et~al.}(2021){van der Wel}, {Bezanson}, {D'Eugenio},
  {Straatman}, {Franx}, {van Houdt}, {Maseda}, {Gallazzi}, {Wu}, {Pacifici},
  {Barisic}, {Brammer}, {Munoz-Mateos}, {Vervalcke}, {Zibetti}, {Sobral}, {de
  Graaff}, {Calhau}, {Kaushal}, {Muzzin}, {Bell}, \& {van
  Dokkum}}]{VanDerWel2021}
{van der Wel}, A., {Bezanson}, R., {D'Eugenio}, F., {et~al.} 2021, \apjs, 256,
  44, \dodoi{10.3847/1538-4365/ac1356}

\bibitem[{{Verrico} {et~al.}(2022){Verrico}, {Setton}, {Bezanson}, {Greene},
  {Suess}, {Goulding}, {Spilker}, {Kriek}, {Feldmann}, {Narayanan}, {Donofrio},
  \& {Khullar}}]{Verrico2022}
{Verrico}, M., {Setton}, D.~J., {Bezanson}, R., {et~al.} 2022, arXiv e-prints,
  arXiv:2211.16532.
\newblock \doarXiv{2211.16532}

\bibitem[{{Weaver} {et~al.}(2022){Weaver}, {Davidzon}, {Toft}, {Ilbert},
  {McCracken}, {Gould}, {Jespersen}, {Steinhardt}, {Lagos}, {Capak}, {Casey},
  {Chartab}, {Faisst}, {Hayward}, {Kartaltepe}, {Kauffmann}, {Koekemoer},
  {Kokorev}, {Laigle}, {Liu}, {Long}, {Magdis}, {McPartland}, {Milvang-Jensen},
  {Mobasher}, {Moneti}, {Peng}, {Sanders}, {Shuntov}, {Sneppen}, {Valentino},
  {Zalesky}, \& {Zamorani}}]{Weaver2022}
{Weaver}, J.~R., {Davidzon}, I., {Toft}, S., {et~al.} 2022, arXiv e-prints,
  arXiv:2212.02512, \dodoi{10.48550/arXiv.2212.02512}

\bibitem[{{Webb} {et~al.}(2020){Webb}, {Balogh}, {Leja}, {van der Burg},
  {Rudnick}, {Muzzin}, {Boak}, {Cerulo}, {Gilbank}, {Lidman}, {Old},
  {Pintos-Castro}, {McGee}, {Shipley}, {Biviano}, {Chan}, {Cooper}, {De Lucia},
  {Demarco}, {Forrest}, {Jablonka}, {Kukstas}, {McCarthy}, {McNab}, {Nantais},
  {Noble}, {Poggianti}, {Reeves}, {Vulcani}, {Wilson}, {Yee}, \&
  {Zaritsky}}]{Webb2020}
{Webb}, K., {Balogh}, M.~L., {Leja}, J., {et~al.} 2020, \mnras, 498, 5317,
  \dodoi{10.1093/mnras/staa2752}

\bibitem[{{Wellons} {et~al.}(2015){Wellons}, {Torrey}, {Ma}, {Rodriguez-Gomez},
  {Vogelsberger}, {Kriek}, {van Dokkum}, {Nelson}, {Genel}, {Pillepich},
  {Springel}, {Sijacki}, {Snyder}, {Nelson}, {Sales}, \&
  {Hernquist}}]{Wellons2015}
{Wellons}, S., {Torrey}, P., {Ma}, C.-P., {et~al.} 2015, \mnras, 449, 361,
  \dodoi{10.1093/mnras/stv303}

\bibitem[{{Whalen} {et~al.}(2022){Whalen}, {Hickox}, {Coil}, {Diamond-Stanic},
  {Geach}, {Moustakas}, {Rudnick}, {Rupke}, {Sell}, {Tremonti}, {Davis},
  {Perrotta}, \& {Petter}}]{Whalen2022}
{Whalen}, K.~E., {Hickox}, R.~C., {Coil}, A.~L., {et~al.} 2022, arXiv e-prints,
  arXiv:2209.13632.
\newblock \doarXiv{2209.13632}

\bibitem[{{Whitaker} {et~al.}(2012){Whitaker}, {Kriek}, {van Dokkum},
  {Bezanson}, {Brammer}, {Franx}, \& {Labb{\'e}}}]{Whitaker2012a}
{Whitaker}, K.~E., {Kriek}, M., {van Dokkum}, P.~G., {et~al.} 2012, \apj, 745,
  179, \dodoi{10.1088/0004-637X/745/2/179}

\bibitem[{{Wild} {et~al.}(2016){Wild}, {Almaini}, {Dunlop}, {Simpson},
  {Rowlands}, {Bowler}, {Maltby}, \& {McLure}}]{Wild2016}
{Wild}, V., {Almaini}, O., {Dunlop}, J., {et~al.} 2016, \mnras, 463, 832,
  \dodoi{10.1093/mnras/stw1996}

\bibitem[{{Wild} {et~al.}(2020){Wild}, {Taj Aldeen}, {Carnall}, {Maltby},
  {Almaini}, {Werle}, {Wilkinson}, {Rowlands}, {Bolzonella}, {Castellano},
  {Gargiulo}, {McLure}, {Pentericci}, \& {Pozzetti}}]{Wild2020}
{Wild}, V., {Taj Aldeen}, L., {Carnall}, A., {et~al.} 2020, \mnras, 494, 529,
  \dodoi{10.1093/mnras/staa674}

\bibitem[{{Worthey} \& {Ottaviani}(1997)}]{Worthey1997}
{Worthey}, G., \& {Ottaviani}, D.~L. 1997, \apjs, 111, 377,
  \dodoi{10.1086/313021}

\bibitem[{{Wu} {et~al.}(2018){Wu}, {van der Wel}, {Bezanson}, {Gallazzi},
  {Pacifici}, {Straatman}, {Bari{\v{s}}i{\'c}}, {Bell}, {Chauke}, {van Houdt},
  {Franx}, {Muzzin}, {Sobral}, \& {Wild}}]{Wu2018}
{Wu}, P.-F., {van der Wel}, A., {Bezanson}, R., {et~al.} 2018, \apj, 868, 37,
  \dodoi{10.3847/1538-4357/aae822}

\bibitem[{{Yesuf}(2022)}]{Yesuf2022}
{Yesuf}, H.~M. 2022, arXiv e-prints, arXiv:2207.12844.
\newblock \doarXiv{2207.12844}

\bibitem[{{Zabludoff} {et~al.}(1996){Zabludoff}, {Zaritsky}, {Lin}, {Tucker},
  {Hashimoto}, {Shectman}, {Oemler}, \& {Kirshner}}]{Zabludoff1996}
{Zabludoff}, A.~I., {Zaritsky}, D., {Lin}, H., {et~al.} 1996, \apj, 466, 104,
  \dodoi{10.1086/177495}

\bibitem[{{Zhou} {et~al.}(2020){Zhou}, {Newman}, {Dawson}, {Eisenstein},
  {Brooks}, {Dey}, {Dey}, {Duan}, {Eftekharzadeh}, {Gazta{\~n}aga}, {Kehoe},
  {Landriau}, {Levi}, {Licquia}, {Meisner}, {Moustakas}, {Myers},
  {Palanque-Delabrouille}, {Poppett}, {Prada}, {Raichoor}, {Schlegel},
  {Schubnell}, {Staten}, {Tarl{\'e}}, \& {Y{\`e}che}}]{Zhou2020}
{Zhou}, R., {Newman}, J.~A., {Dawson}, K.~S., {et~al.} 2020, Research Notes of
  the American Astronomical Society, 4, 181, \dodoi{10.3847/2515-5172/abc0f4}

\bibitem[{{Zhou} {et~al.}(2022){Zhou}, {Dey}, {Newman}, {Eisenstein}, {Dawson},
  {Bailey}, {Berti}, {Guy}, {Lan}, {Zou}, {Aguilar}, {Ahlen}, {Alam}, {Brooks},
  {de la Macorra}, {Dey}, {Dhungana}, {Fanning}, {Font-Ribera}, {Gontcho},
  {Honscheid}, {Ishak}, {Kisner}, {Kov{\'a}cs}, {Kremin}, {Landriau}, {Levi},
  {Magneville}, {Martini}, {Meisner}, {Miquel}, {Moustakas}, {Myers}, {Nie},
  {Palanque-Delabrouille}, {Percival}, {Poppett}, {Prada}, {Raichoor}, {Ross},
  {Schlafly}, {Schlegel}, {Schubnell}, {Tarl{\'e}}, {Weaver}, {Wechsler},
  {Y{\`e}che}, \& {Zhou}}]{Zhou2022}
{Zhou}, R., {Dey}, B., {Newman}, J.~A., {et~al.} 2022, arXiv e-prints,
  arXiv:2208.08515.
\newblock \doarXiv{2208.08515}

\bibitem[{{Zou} {et~al.}(2017){Zou}, {Zhou}, {Fan}, {Zhang}, {Zhou}, {Nie},
  {Peng}, {McGreer}, {Jiang}, {Dey}, {Fan}, {He}, {Jiang}, {Lang}, {Lesser},
  {Ma}, {Mao}, {Schlegel}, \& {Wang}}]{Zou2017}
{Zou}, H., {Zhou}, X., {Fan}, X., {et~al.} 2017, \pasp, 129, 064101,
  \dodoi{10.1088/1538-3873/aa65ba}

\end{thebibliography}

\end{document}